\documentclass[jgr]{agutex}

\usepackage[dvips]{graphicx}
\usepackage{verbatim}
\usepackage{amsmath}
\usepackage{amssymb}
\usepackage{multirow}
\usepackage{url}
\usepackage{color}
\usepackage{ulem}
\usepackage[figuresleft]{rotating}
\newcommand{\ve}[1]{\mathbf{#1}}

\paperid{2016JA023075}
\authorrunninghead{WANG ET AL.}
\titlerunninghead{Twists of Interplanetary MFRs}

\begin{document}

\title{On the twists of interplanetary magnetic flux ropes observed at 1 AU}

\author{Yuming Wang,\altaffilmark{1,3,*} Bin Zhuang,\altaffilmark{1,4} Qiang Hu,\altaffilmark{2} Rui Liu,\altaffilmark{1,4,5} Chenglong Shen,\altaffilmark{1,3,5} and Yutian Chi\altaffilmark{1,4}}

\affil{$^1$ CAS Key Laboratory of Geospace Environment, Department
of Geophysics and Planetary Sciences, University of Science and
Technology of China, Hefei, Anhui 230026, China}

\affil{$^2$ Department of Space Science and CSPAR, The University of Alabama in Huntsville, Huntsville, Alabama, USA}


\affil{$^3$ Synergetic Innovation Center of Quantum Information and
Quantum Physics, University of Science and Technology of China,
Hefei, Anhui 230026, China}

\affil{$^4$ Collaborative Innovation Center of Astronautical Science and Technology, China}

\affil{$^5$ Mengcheng National Geophysical Observatory, School of
Earth and Space Sciences, University of Science and Technology of
China, Hefei, China}

\affil{$^*$ Corresponding Author, Contact: ymwang@ustc.edu.cn}

\begin{abstract}
Magnetic flux ropes (MFRs) are one kind of fundamental structures in the solar/space physics, and
involved in various eruption phenomena. Twist, characterizing how the magnetic field lines wind
around a main axis, is an intrinsic property of MFRs, closely related to the magnetic free energy
and stableness. Although the effect of the twist on the behavior of MFRs had been widely studied in observations,
theory, modeling and numerical simulations, it is still unclear how much amount of twist is carried by
MFRs in the solar atmosphere and in heliosphere and what role the twist played in the eruptions of MFRs.
Contrasting to the solar MFRs, there are lots of in-situ measurements of magnetic clouds (MCs),
the large-scale MFRs in interplanetary space, providing some important information of the twist of MFRs.
Thus, starting from MCs, we investigate the twist of interplanetary MFRs with the aid of a velocity-modified
uniform-twist force-free flux rope model. It
is found that most of MCs can be roughly fitted by the model and nearly half of them
can be fitted fairly well though the derived twist is probably over-estimated by a factor of 2.5.
By applying the model to 115 MCs observed at 1 AU, we find that (1) the twist angles of interplanetary MFRs
generally follow a trend of about $0.6\frac{l}{R}$ radians, where $\frac{l}{R}$ is the aspect ratio of a MFR,
with a cutoff at about $12\pi$ radians AU$^{-1}$,
(2) most of them are significantly larger than
$2.5\pi$ radians but well bounded by $2\frac{l}{R}$ radians, (3) strongly twisted magnetic
field lines probably limit the expansion and size of MFRs, and (4) the magnetic field lines in the legs
wind more tightly than those in the leading part of MFRs. These results not only advance
our understanding of the properties and behavior of interplanetary MFRs, but also shed light on
the formation and eruption of MFRs in the solar atmosphere. A discussion about the twist and
stableness of solar MFRs are therefore given.
\end{abstract}

\begin{article}

\section{Introduction}
Magnetic flux ropes (MFRs) are one of the fundamental structures in plasma physics, space physics and astrophysics,
and may exist in different scales from as small as formed in reconnection regions to as large as appeared in
astrophysical jets. MFRs can be defined when a bunch of magnetic field lines demonstrate a
systematic and significant {\it twist} around an internal main axis. In mathematics, the quantity,
{\it twist} (in units of radians per unit length), is described as $T=\frac{B_{\varphi}}{rB_z}$ in local cylindrical coordinates $(r, \varphi, z)$
with the $z$-axis along the main axis. It is an important parameter characterizing a MFR. A
strong twisted MFR carries more magnetic free energy density than a weak twisted MFR, and may be subject
to various instabilities.

In solar physics, kink instability is one of the most common instabilities, frequently observed
during solar eruptions~\citep[e.g.,][]{Rust_Kumar_1996, DeVore_Antiochos_2000, Ji_etal_2003,
Williams_etal_2005, Rust_LaBonte_2005}.
Lots of theoretical and numerical simulation studies had shown that a MFR becomes unstable when the
twist exceeds a critical value~\citep[e.g.,][]{Dungey_Loughhead_1954, Kruskal_etal_1958, Hood_Priest_1979, Mikic_etal_1990,
Baty_2001, Fan_Gibson_2004,Torok_Kliem_2005}.
A well-known critical twist is that derived by~\citet{Hood_Priest_1981} for a line-tying force-free MFR with the
uniform-twist solution first proposed by \citet[][called GH model hereafter and see Sec.\ref{sec_model} for
the solution; For clarification, most acronyms and symbols used in this paper are summarized in Appendix A and Table~\ref{tb_par}]{Gold_Hoyle_1960}. They found that the MFR will become
kink unstable when the total twist angle, $\Phi_T$, exceeds $2.5\pi$ radians or the total number
of turns exceeds 1.25 (hereafter called HP critical twist). Here the total twist angle
is the angle of the magnetic field lines rotating around the main axis from one end of the MFR to the other
given by $\int_0^l Tdz$ where $l$ is the length of the main axis.

Actually, the value of critical twist depends on many factors, including the internal magnetic field
configuration~\citep{Dungey_Loughhead_1954, Hood_Priest_1979, Mikic_etal_1990, Bennett_etal_1999, Baty_2001},
the external field~\citep{Hood_Priest_1980, Bennett_etal_1999,
Torok_Kliem_2005}, the plasma $\beta$~\citep{Hood_Priest_1979}, the axial plasma flow~\citep{Zaqarashvili_etal_2010}, etc.
For example, some previous studies~\citep{Dungey_Loughhead_1954, Hood_Priest_1979, Bennett_etal_1999, Baty_2001}
demonstrated that the critical total twist angle, $\Phi_{c}$, is a function
of the aspect ratio (the ratio of the axial length $l$ to the radius $R$) of a MFR, i.e.,
\begin{eqnarray}
\Phi_c=\omega_c\frac{l}{R}\label{eq_critical}
\end{eqnarray}
where $\omega_c$ is a parameter depending on detailed configuration of the MFR. For
another type of uniform-twist flux rope~\citep[first proposed by][]{Alfven_1950},
which has the uniform axial magnetic field and is in a non-force-free state, \citet{Dungey_Loughhead_1954}
and \citet{Bennett_etal_1999} found that $\omega_c$ is about 2, suggesting that a thin MFR has a higher critical twist.
Similar dependence was also investigated by~\citet{Hood_Priest_1979} and \citet{Baty_2001} for various types of flux ropes,
in which $\omega_c$ varies in a large range.
The core structures of solar coronal mass ejections (CMEs), the
largest eruptive phenomenon on the Sun, are believed to be MFRs, which form and
develop before and/or during the eruptions in the corona and evolve into interplanetary space. Thus
learning how strong the twist is in MFRs is extremely useful in understanding the eruption and dynamic
evolution of CMEs.

\subsection{Twist of solar MFRs}\label{sec_smfr}
So far there is no mean to directly observe MFRs on the Sun. All the information of the MFRs on the Sun
are obtained indirectly from multi-wavelength observations and modeling studies.
One of the earliest attempts of measuring the twists of solar MFRs was done to prominences~\citep[e.g.,][]{Vrsnak_etal_1991,
Vrsnak_etal_1993}, which were thought to be a good tracer of MFRs. \citet{Vrsnak_etal_1991} analyzed a set
of 28 prominences observed in $H\alpha$ passband with the focus on the helical-shaped threads in the prominences.
By assuming a reasonable flux rope model and that the $H\alpha$ material is frozen-in the magnetic field lines,
they measured the pitch angle of these threads and found that the total twist angles varied in a range roughly
from $5\pi$ to $15\pi$. Since the resolution of the $H\alpha$ images was not good enough at that time, the results
may suffer from large uncertainties. With higher-resolution imaging data, \citet{Romano_etal_2003} investigated
a prominence eruption. By using the same method, they derived that the total twist angle of one helical thread
of the prominence was about $10\pi$ and decreased to about $2\pi$ during the eruption.

More recently, with even higher-resolution imaging data, \citet{Srivastava_etal_2010}
successfully measured the twist of a coronal loop in active region (AR) 10960, which showed
bright-dark alternating streaks along the long axis of a loop in the TRACE 171\AA\ images, implying a highly
twisted structure. By combining the observations from SOHO/MDI,
Hinode/SOT and TRACE, the authors figured out that the aspect ratio of the loop was about 20 and the
total twist angle of the loop was about $12\pi$, and suggested that the kink instability was responsible for a small flare in the AR.
Another similar case could be found in the study of the 2002 July 15 flare by \citet{Gary_Moore_2004}, in which
an erupting four-turn helical structure was clearly observed in the TRACE 1600\AA\ images.
All of these measured twists significantly exceeded the HP critical twist, but might support the other theoretical
studies aforementioned that thin MFRs have higher critical twists for the kink instability~\citep{Dungey_Loughhead_1954, Bennett_etal_1999,
Hood_Priest_1979} as those observed structures did have large aspect ratios.

More efforts on the twists of solar MFRs are from modeling methods. With the aid of a non-linear force-free field
(NLFFF) extrapolation technique, for example, \citet{Yan_etal_2001} presented a MFR above the polarity inversion
line associated with an X5.7-class flare on 2000 July 14. They estimated that the total twist angle of the MFR
was about $3\pi$, and was maintained for about 10 hours before the flare. Similar studies could be found in, e.g.,
\citet{Regnier_etal_2002} and \citet{Guo_etal_2010}, in which they roughly estimated that the twist of MFRs varied from
about $2\pi$ to $3\pi$.
More precisely, \citet{Berger_Prior_2006} gave a general equation (Eq.12 in their paper) for the twist of a bunch of smooth
non-self-intersecting magnetic field lines. It was found that the total twist angle of a magnetic field line for
force-free fields can be approximated as $\frac{\alpha}{2}l_{line}$ (see Eq.16 in \citealt{Berger_Prior_2006} or
Eq.7 in \citealt{LiuR_etal_2016}), where $\alpha$ is the force-free parameter and $l_{line}$ is the length of the field line.
This method was later applied to the extrapolated three-dimensional (3D) magnetic field lines to
infer the twists of candidate MFRs~\citep[e.g.,][]{Inoue_etal_2011, Inoue_etal_2012, Guo_etal_2013,
Chintzoglou_etal_2015, LiuR_etal_2016}.
For example, \citet{Inoue_etal_2011} studied the magnetic field structure surrounding the sheared flare ribbons of an
X3.4-class flare on 2006 December 13, and inferred that the total twist angle varied from about $0.5\pi$ to $1.2\pi$.
\citet{LiuR_etal_2016} investigated the MFRs associated with a series of flares in AR 11817, and found that all of the
MFRs had a moderate twist angle less than $4\pi$. Particularly, from the twist maps in their paper, one may find that
the distribution of the twist was more or less flattened in the MFRs, implying a configuration closer to a uniform-twist
magnetic field structure. Besides, it should be noted that the inferred
twist angle by the \citet{Berger_Prior_2006} equation is not exactly equal to the traditional twist angle, $\Phi_T$,
defined at the beginning of the paper. It is very close to the traditional twist near the axis but deviates at other
places~\citep[see Appendix C of][]{LiuR_etal_2016}, and
should be treated as a local twist angle, labeled $\Phi_L$, contrasting to $\Phi_T$. As can be
seen from Eq.\ref{eq_alpha} and \ref{eq_length} below, for a uniform-twist flux rope with the GH model, there is
$\frac{\Phi_T}{\Phi_L}=\sqrt{1+T^2r^2}$, suggesting an under-estimation of $\Phi_T$.

An interesting thing here is that the inferred MFRs from NLFFF extrapolations have much less twists
than observed helical structures. There are two possible reasons. One is that the extrapolated twists are significantly
under-estimated as demonstrated above with the GH model; the other is that the observed
helical structures might not fully reflect the real twist of magnetic field lines. It is difficult to judge which one is
the case without direct detection of MFRs. Thus, it becomes necessary to investigate the twists
of interplanetary MFRs, most of which are believed to be evolved from the ejected MFRs on the Sun and may be directly
measured by in-situ instruments.

\subsection{Twist of interplanetary MFRs}
The large-scale MFRs in interplanetary space are usually termed magnetic clouds (MCs)~\citep{Burlaga_etal_1981}, a subset
of CMEs. The twists of magnetic field lines inside MCs can be estimated by using
energetic particles released during impulsive flares magnetically connecting to the in-situ detector.
It was often observed that energetic particles
demonstrate a velocity dispersion in the energy-time plot~\citep[e.g.,][]{Kutchko_etal_1982}.
This can be used to infer the lengths of magnetic field lines based on the facts that
energetic particles are fewly scattered during the propagation and particles with higher speeds will arrive
earlier when they are injected into interplanetary space at the same time~\citep{Larson_etal_1997, Mazur_etal_2000,
Kahler_Ragot_2006, Chollet_etal_2007, Ragot_Kahler_2008, Kahler_etal_2011, Kahler_etal_2011a, Tan_etal_2012}.
The desired energy range is above 1 keV for electrons or 20 keV for ions. Electrons are better than
ions because electrons have smaller gyroradii. As long as the length of the MC's main axis can
be determined or reasonably assumed, the twists of the magnetic field lines inside the MC can be deduced from the length.

Concretely speaking, by assuming that the energetic electrons released almost the same time and propagate
along the same bunch of magnetic field lines, the velocity dispersion at in-situ detector can be fitted by
the equation $l_{line}=v(t-t_0)$, where $v$ and $t$ are observed velocities and arrival times, to obtain
the release time, $t_0$, and the field line length $l_{line}$. However, this method often gets $l_{line}<1.2$ AU,
less than the typical length of Parker spiral field lines, due to the large uncertainty in the
measurements~\citep{Kahler_Ragot_2006}. Thus alternatively, people used the onset of the associated Type III radio burst as the release
time of the received energetic electrons, and derived the field line length based on the same equation with
the free parameter, $t_0$, fixed. By using this method,~\citet{Larson_etal_1997} inferred that the magnetic field
line length varied from about 3 AU near the edge to about 1.2 AU near the center of the magnetic cloud
detected by the Wind spacecraft during 1995 October 18--20. By using the electrons with higher energy than those
used in~\citet{Larson_etal_1997}, ~\citet{Kahler_etal_2011} got a similar result. Further, they
expanded the study to more MC events, and found that the field line lengths inside the MCs are ranged between
about 1.3 and 3.7 AU. It can be roughly inferred from the length range by using Eq.\ref{eq_length} below that the twists of the magnetic field
lines inside MCs actually do not vary too much. These inferred lengths are notably deviated from those predicted by the flux rope model
with Lundquist solution~\citep{Lundquist_1950}, in which the field line length becomes infinitely large
when approaching the edge of a MFR.

In addition to the probes of energetic particles, Grad-Shafranov (GS) reconstruction technique is another
approach to infer the twist of MCs. Different from other MC's flux rope models, it does not preset any magnetic
configuration of the MFR, and instead it can infer the magnetic field vector in the plane perpendicular to the MFR axis under
the magnetohydrostatic assumption~\citep{Hu_Sonnerup_2002}. By assuming $\frac{\partial}{\partial z}=0$ with
$\hat{\ve z}$ along the axis, \citet{Hu_etal_2014} drew out magnetic field lines from the plane for 18 MCs of
interest and studied the twists inside the MCs. They found that the twist changes in a small range from the
axis to the edge for most events, and the average twist or the number of turns per unit length, $\tau$ (refer to
Eq.\ref{eq_tau} in the next section for its definition), varies
between $\sim1.7$ to $\sim7.7$ turns per AU with one exceptional large $\tau$ of about $15$ turns per AU.
A similar case study can be found in an earlier paper by \citet{Mostl_etal_2009}, in which the authors used
multi-spacecraft measurements to reconstruct a MC and inferred a twist of about 1.5 -- 1.7 turns per AU.
Further,~\citet{Hu_etal_2015}
compared the deduced magnetic field line lengths with those estimated from the energetic electrons by \citet{Kahler_etal_2011},
and a good correlation was found. Since the flat change in twist from the axis to the edge of the MCs was found
through both electron probes and GS reconstructions, \citet{Hu_etal_2015}
also argued that the magnetic field lines of MCs are more likely to be uniformly twisted, and therefore the
uniform-twist flux rope with the GH solution,
rather than the Lundquist flux rope or others with a highly non-uniform twist should be used to model
the interplanetary MFRs.
These results are quite consistent with the studies of solar MFRs as introduced in Sec.\ref{sec_smfr}.

Although \citet{Hu_etal_2015} mentioned and applied the GH model, they only used it to
estimate the magnetic field line length based on GS fitting results. A full application of the GH model
in fitting of interplanetary MFRs was rarely reported. To our knowledge, the first study in such kind
is that by \citet{Farrugia_etal_1999}. They investigated a MFR during 1995 October 24--25 observed by Wind,
and inferred that the twist of the magnetic field lines in the MFR was about 8 turns per AU.
\citet{Dasso_etal_2006} also applied the GH model to study the helicity and fluxes of the MC
on 1995 October 18--20. It was suggested that the twist of the MC is about 2.4 turns per AU based on their
GH model. As will see below, this value is quite consistent with the inferred magnetic field line
lengths~\citep{Kahler_etal_2011} by assuming an axial length of about 2.57 AU.

Inspired by the studies of the twists of solar and interplanetary MFRs, in this paper we try to apply
the GH model to a large sample of interplanetary MCs, check how many and how well interplanetary
MCs can be fitted by the model, and seek some statistical properties of MFRs in terms of twist.
First, we develop the original GH model into the velocity-modified GH model, which will be introduced
in detail in the next section. Secondly in Section~\ref{sec_compare}, by applying the model to some MC events, we compare the deduced magnetic field
line lengths and twists to those reported in \citet{Kahler_etal_2011} and \citet{Hu_etal_2015}, respectively,
to justify the model. A statistical analysis of the twists of MCs are then presented in Sec.\ref{sec_statistics}.
We believe that the method established in this paper and the results obtained
will be a useful complement to the currently existing approaches and results for interplanetary MFRs and
also helpful to understand the properties and behaviors of MFRs on the Sun.

\section{Velocity-modified uniform-twist flux rope model with the GH solution}
\subsection{Description of the model}\label{sec_model}

The reason we incorporate
velocity into the model is that MCs are dynamically evolving and the measurements of
in-situ 3D velocity may provide additional
constraints on the fitting procedure.
The derivation of the model is similar to that of the velocity-modified
cylindrical flux rope model with Lundquist solution by~\citet{Wang_etal_2015}. The main
difference is that we here replace the Lundquist solution of the magnetic field with
the GH solution. The former is linear force-free and the latter is non-linear force-free.
For the completeness and clarification, the model and how to
evaluate the goodness-of-fit are briefly described below.

\begin{figure*}[tb]
  \centering
  \includegraphics[width=\hsize]{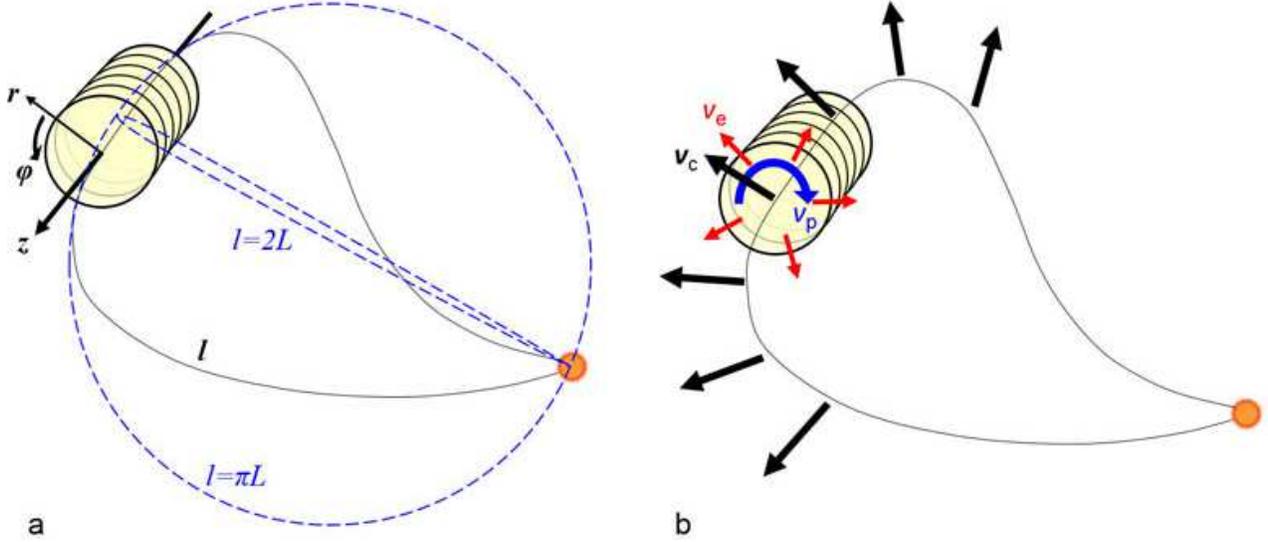}
  \caption{(a) Schematic picture of an MC at the heliocentric distance of $L$ (adapted from \citet{Wang_etal_2009}).
The black line indicates the looped axis of the MC with a length of $l$. The blue dashed lines suggest the upper
and lower limits of $l$. (b) Illustration of the three types of global motions of an MC. The black, red and blue
arrows denote the linear propagating motion, expanding motion and poloidal motion, respectively.}\label{fg_coord}
\end{figure*}

We consider a loop-like global geometry of the interplanetary MFR as shown in Figure~\ref{fg_coord},
and investigate a segment of the MFR in the cylindrical coordinates $(r,\varphi,z)$. The
coordinates and symbols used in this study are exactly the same as those in~\citet{Wang_etal_2015}.
The uniform-twist magnetic field inside the MFR is described as
\begin{eqnarray}
&&B_r=0\\
&&B_\varphi=\frac{Tr}{1+T^2r^2}B_0\label{eq_b2i}\\
&&B_z=\frac{1}{1+T^2r^2}B_0\label{eq_b3i}
\end{eqnarray}
in which $B_0$ is the magnetic field at the MFR axis where $r=0$ and $T$ is the twist of the magnetic field
lines in units of radians per unit length as defined in Introduction.
A positive/negative value of $T$ means the handedness of the MFR is right/left.
Let the length of the MFR's axis be $l$, the number of turns of the field lines winding around
the axis from one end of the MFR to the other is given by
\begin{eqnarray}
n=\frac{T}{2\pi}l=\frac{\Phi_T}{2\pi}\label{eq_twist}
\end{eqnarray}
or the number of turns per unit length along the MFR axis by
\begin{eqnarray}
\tau=\frac{T}{2\pi}\label{eq_tau}
\end{eqnarray}
By assuming the self-similar evolution of the MFR, $l$ can be given by
\begin{eqnarray}
l=\lambda L\label{eq_l}
\end{eqnarray}
where $L$ is the heliocentric distance of the leading part of the MC as illustrated in Figure~\ref{fg_coord}, and $\lambda$
is a constant, named effective length factor here.
The axial length $l$ or the effective length factor $\lambda$ is of importance to estimate the total magnetic flux, helicity and
magnetic energy carried by the MFR. Here, we set $\lambda$ to be
$\frac{\pi+2}{2}\pm\frac{\pi-2}{2}\approx2.57\pm0.57$ following \citet{Wang_etal_2015} and summarized in Figure~\ref{fg_coord}a,
which is almost same to the value of $2.6\pm0.3 AU$ inferred by~\citet{Demoulin_etal_2016} and also
very close to $2.7\pm0.5$ assumed by~\citet{Kahler_etal_2011}.
By comparing the directly probed magnetic field line lengths from the energy electrons and the twists derived
from GS model, \citet{Hu_etal_2015} concluded that the effective axial length, $L_{eff}$, from the footpoint on the
Sun to the MC observed at 1 AU is ranged between 1 to 2 AU, corresponding to a $\lambda$ from 2 to 4, which is slightly
wider than the range of $\lambda$ we used here.

Further, the non-constant force-free parameter $\alpha$ is given by
\begin{eqnarray}
\alpha=\frac{2T}{1+T^2r^2}\label{eq_alpha}
\end{eqnarray}
the length of the magnetic field line on any torus by
\begin{eqnarray}
l_{line}=\int_0^{l}\sqrt{dz^2+(rd\varphi)^2}=l\sqrt{1+T^2r^2}\label{eq_length}
\end{eqnarray}
the axial and poloidal magnetic fluxes by
\begin{eqnarray}
&&F_z=\int_0^{2\pi}\int_0^R B_zrdrd\varphi=\frac{B_0l^2}{4\pi n^2}\ln\left(1+4\pi^2n^2\frac{R^2}{l^2}\right)\label{eq_flux_z} \\
&&F_\varphi=\int_0^l\int_0^R|B_\varphi| drdz=\frac{B_0l^2}{4\pi |n|}\ln\left(1+4\pi^2n^2\frac{R^2}{l^2}\right)\label{eq_flux_p}
\end{eqnarray}
where $R$ is the radius of the MFR, and the total magnetic helicity by~\citep{Berger_Field_1984}
\begin{eqnarray}
H_m=nF_z^2\label{eq_helicity}
\end{eqnarray}

It is interesting to discuss the invariance of some parameters. For a perfectly conducting plasma in a closed volume,
the total magnetic helicity is constant, which is believed to be a good approximation for MCs.
Together with the assumption of constant axial magnetic flux, one may infer from \ref{eq_helicity}
that $n$ is invariant, and the self-similar assumption used in this model implies that $\frac{R}{l}$ is invariant.
The invariance of $n$ could be illuminated from
another angle of view. One may imagine that for a given magnetic field line in a flux rope with finite length,
its value of $n$ is related to the positions of the fluid elements frozen onto the two ends of the field lines. These fluid
elements are supposed to locate on the surface of the Sun based on the picture that a magnetic cloud is a
looped structure with two ends rooted on the Sun~\citep{Kahler_Reames_1991, Larson_etal_1997}. As long as
the relative positions of the fluid elements do not change significantly during the MFR passing through
the in-situ observer, the configuration of the magnetic field lines as well as $n$ can be treated unchanged.
Then, as a consequence, the parameters $T$, $\tau$ and $\alpha$ are time- or distance-dependent. Concretely,
the twist of MCs decreases in such a way to keep $n$ being constant when they propagate and expand into interplanetary space.

Further, we may infer that
\begin{eqnarray}
B_0\propto L^{-2}
\end{eqnarray}
and the magnetic energy
\begin{eqnarray}
E_m&=&\int_0^l\int_0^{2\pi}\int_0^R\frac{B^2}{2\mu}rdrd\varphi dz=\frac{B_0^2l^3}{8\pi\mu n^2}\ln\left(1+4\pi^2n^2\frac{R^2}{l^2}\right) \nonumber \\
&\propto & L^{-1}\label{eq_energy}
\end{eqnarray}
These scaling laws are the same as those for the Lundquist flux ropes.

By defining a dimensionless parameter, $x=\frac{r}{R}$, to be the normalized distance from the axis of the MFR,
we can find that the twist, $T$, and the radius, $R$, cannot be distinguished in the GH solution (Eq.\ref{eq_b2i} and \ref{eq_b3i}).
In the Lundquist solution, the radius, i.e., the boundary, of a MFR is usually set at the first zero of the zero-order
Bessel function $J_0(r)$ where $B_z$ vanishes. However, the GH solution does not have such a
zero point along the $r$-axis. One potential special point locates at $r=\frac{1}{T}$, where $B_\varphi$ reaches the maximum.
This point was assumed to be the boundary of the MFR in a series of papers by \citeauthor{Hood_Priest_1979} in, e.g., 1979, 1980 and 1981.

\begin{table*}
\linespread{1.5} \caption{Parameters involved in the
velocity-modified GH model}\label{tb_par}
\begin{center}
\begin{tabular}{c|p{430px}}
\hline Parameter & Explanation \\
\hline
\multicolumn{2}{c}{Free parameters in the model} \\
\hline
$B_0(t)$ & Magnetic field strength at the axis of the MFR.\\
$\omega$ & A parameter containing the information of the twist (Eq.\ref{eq_critical} or \ref{eq_r-omega}).\\
$\theta$ & Elevation angle of the axis of the MFR in GSE.\\
$\phi$ & Azimuthal angle of the axis of the MFR in GSE.\\
$d$ & The closest approach of the observational path to the axis of
the MFR.\\
$v_X$ & Propagation speed of the MFR in the direction of $\mathbf{\hat{X}}$.\\
$v_Y$ & Propagation speed of the MFR in the direction of $\mathbf{\hat{Y}}$.\\
$v_Z$ & Propagation speed of the MFR in the direction of $\mathbf{\hat{Z}}$.\\
$v_{e}$ & Expansion speed of the boundary of the MFR in the direction of $\mathbf{\hat{r}}$.\\
$v_{p}(t)$ & Poloidal speed at the boundary of the MFR in the direction of $\mathbf{\hat{\varphi}}$.\\
\hline
\multicolumn{2}{c}{Other derived parameters from the model} \\
\hline
$R(t)$ & Radius of the cross-section of the MFR.\\
$t_c$ & The time when the observer arrives at the closest approach.\\
$\Theta$ & Angle between the axis of the MFR and $\mathbf{\hat{X}}$-axis.\\
$\alpha(t)$ & Non-constant force-free parameter (Eq.\ref{eq_alpha}).\\
$l_{line}(t)$ & Lengths of the magnetic field lines from one end of the MFR to the other (Eq.\ref{eq_length}).\\
$T(t)$ & Twist per unit length along the MFR axis.\\
$\tau(t)$ & Number of the turns per unit length along the MFR axis, i.e., $\frac{T}{2\pi}$.\\
$\Phi_T$ & Total twist angle, i.e., integration of $T$ along the MFR's axis from one end to the other.
$\Phi_c$ and $\Phi_L$ refer to the critical and local total twist angles, respectively.\\
$n$ & Total number of turns of the magnetic field lines of the MFR, i.e., $\frac{\Phi_T}{2\pi}$.\\
$F_z$ & Axial magnetic flux of the MFR (Eq.\ref{eq_flux_z}).\\
$F_\varphi$ & Poloidal magnetic flux of the MFR (Eq.\ref{eq_flux_p}).\\
$H_m$ & Total magnetic helicity of the MFR (Eq.\ref{eq_helicity}).\\
$E_{m}(t)$ & Total magnetic energy of the MFR (Eq.\ref{eq_energy}).\\
$\chi_n$ & Normalized root mean square (rms) of the difference between the
modeled results and observations (Eq.\ref{eq_chi}).\\
\hline
\end{tabular}
\end{center}
\end{table*}

Actually, the boundary of a GH flux rope can be freely chosen.
We may introduce a new parameter, $\omega$, to relate $R$ with $T$ as follows
\begin{eqnarray}
R=\frac{\omega}{T}\label{eq_r-omega}
\end{eqnarray}
where $\omega$ could be any non-zero value. It should be noted that $\omega=RT=\frac{R}{l}\Phi_T$ has the same dimension
as $\omega_c$ in Eq.\ref{eq_critical}, suggesting that searching the critical twist angle $\Phi_T$ is equivalent to searching
the critical value of $\omega$ for a GH flux rope. Since $\frac{R}{l}$ and $\Phi_T=2\pi n$ are constant as discussed above, $\omega$
is also constant and therefore time- or distance-independent.
Equations \ref{eq_b2i} and \ref{eq_b3i} are then rewritten as
\begin{eqnarray}
&&B_\varphi=\frac{\omega x}{1+\omega^2x^2}B_0\label{eq_b2}\\
&&B_z=\frac{1}{1+\omega^2x^2}B_0\label{eq_b3}
\end{eqnarray}
The determination of the value of $\omega$ is important, and will be introduced in the next section.

The three components of the global plasma motion of the MFR, which are the propagation motion,
expanding motion and the poloidal motion (ref. to Fig.\ref{fg_coord}b), are respectively given as
\begin{eqnarray}
&&\ve v_c=(v_X,v_Y,v_Z) \label{eq_vc}\\
&&v_r(x)=xv_e \label{eq_ve}\\
&&v_\varphi(t,x)=v_p(t)=v_p(t_0)\frac{R(t_0)}{R(t)}\label{eq_vp}
\end{eqnarray}
where $v_e$ and $v_p$ are the expanding and poloidal speeds at the boundary of the MFR, respectively,
and $t_0$ is a reference time. Here, the propagation speed, $\ve v_c$, and the expansion speed, $v_e$, are
assumed constant during the passage of the MC. The expansion speed and poloidal speed given by Eq.\ref{eq_ve} and \ref{eq_vp}
are designed to satisfy the self-similar evolution assumption and the latter also satisfies the mass and angular momentum
conservations~\citep[see Sec.2.1.1 of][]{Wang_etal_2015}.
Consequently, we have
\begin{eqnarray}
&&R(t)=R(t_0)+v_e(t-t_0)\\
&&B_0(t)=B_0(t_0)\left[\frac{R(t_0)}{R(t)}\right]^2\label{eq_b0}
\end{eqnarray}
The latter is required by the magnetic flux conservation, and these two equations connect the equations of velocities
(Eq.\ref{eq_vc}--\ref{eq_vp}) with the equations of the magnetic field structure (Eq.\ref{eq_b2}--\ref{eq_b3}
or Eq.\ref{eq_b2i}--\ref{eq_b3i}).
Equations \ref{eq_b2}--\ref{eq_b0} form the model, in which a total of 10 free parameters, listed
in Table~\ref{tb_par}, need to be determined by fitting the model to the measurements
of the magnetic field and velocity.
Some derivable parameters are also listed in the table,
in which the values of $R$, $n$, $l_{line}$, $F_\varphi$, $H_m$ and $E_m$ depends on the axial length $l$.
It should be noted that the radius, $R$, is not a free parameter, because
it can be uniquely determined by the closest approach, $d$, and the propagation velocity, $\ve v_c$.

\subsection{Evaluation of the quality of fit}\label{sec_quality}

Starting from a series of initial values, all of the free parameters except
for $\omega$ are optimized by using a least squares fitting procedure, which
is the exactly same as that described in Sec.2.1.3 of~\citet{Wang_etal_2015}
and will not be repeated here. The parameter $\omega$ is estimated separately by
using
\begin{eqnarray}
\omega=RT=\frac{B_\varphi}{xB_z}\label{eq_omega}
\end{eqnarray}
in which $B_\varphi$ and $B_z$ are the measured magnetic field. This equation depends on
the axial orientation ($\theta$ and $\phi$) and the closest approach ($d$), which will be
changed during the fitting. Thus, we embed the estimation of $\omega$ into the fitting procedure to
make sure that the value of $\omega$ is re-calculated once the orientation and/or the closest approach change.
The reason to make the
estimate of $\omega$ standalone is to provide an
additional condition to constrain the value of $T$ and $R$, which are actually coupled
in the GH model, and also provide a method to evaluate the goodness of the uniform-twist
assumption.

\begin{figure*}[tb]
  \centering
  \includegraphics[width=\hsize]{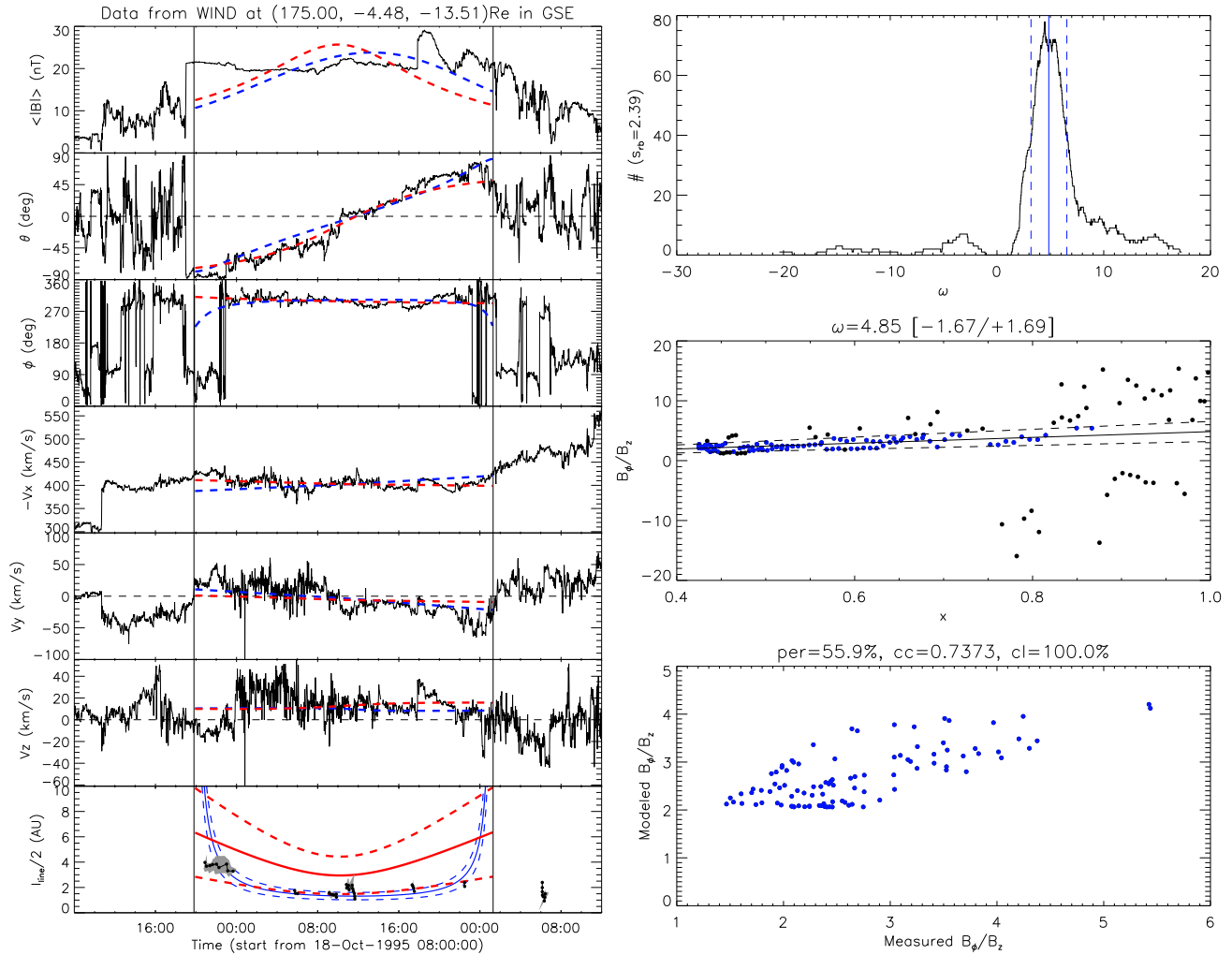}
  \caption{{\it Left column}: Interplanetary magnetic field and solar wind velocity recorded by Wind spacecraft
for the MC No.1. From the top to the bottom, the profiles show the total magnetic field strength,
elevation and azimuthal angles of magnetic field vector, three components of the velocity in GSE coordinates
and the magnetic field line lengths inferred by \citet{Kahler_etal_2011}. The red/blue dashed lines in the first six
panels are the fitting curves of the velocity-modified GH/Lundquist model. The red/blue lines in the last panel are the
magnetic field line length modeled by the corresponding models (the dashed lines indicate uncertainties).
{\it Right column}: The top panel
is the histogram of $\omega$, the middle panel shows the correlation between $\frac{B_\varphi}{B_z}$ and $x$
for all the measurements in the MC, and the bottom panel presents the correlation between the modeled and
measured $\frac{B_\varphi}{B_z}$. See the text in Sec.\ref{sec_quality} for details.}\label{fg_mc001}
\end{figure*}

To illustrate how to estimate the value of
$\omega$, we show an example in Figure~\ref{fg_mc001}. It is a well-known MC observed during 1995
October 18--19, which was investigated by~\citet{Larson_etal_1997} for its global configuration and
by~\citet{Kahler_etal_2011} and~\citet{Hu_etal_2015} for the twists of the magnetic field lines inside
the MC. The boundaries of the MC marked by two solid lines in the left panel of Figure~\ref{fg_mc001} are chosen from
Lepping's list~\citep{Lepping_etal_2006}, which are slightly different from those identified
by~\citet{Larson_etal_1997}.

First, we calculate $\omega_i=\frac{B_{i\varphi}}{x_iB_{iz}}$ for all
the data points in the MC interval based on Eq.\ref{eq_omega}. Due to the presence of the possible
random fluctuations and small-scale features in the measured magnetic field, the values of $\omega_i$ are
probably scattered in a large range though most of them may concentrate around a certain value (as illustrated
in the top two panels on the right column of Fig.\ref{fg_mc001}). Those minor data points near the two ends
of the $\omega_i$ range may bias the estimated value of $\omega$. To reduce the possible bias, we narrow down the range of $\omega_i$
to remove those minor data points until 10\% of the original data points are excluded. In other words, we select the
data points falling in the $\omega_i$ range at the significance of 90\%.

Second, we use a bin running through the range to count how many data points
fall in the bin, and generate a histogram from these counts for further analysis. The step used to move the bin
from one end of the $\omega_i$ range to the other is set to be 0.01, and the size of the running bin is determined by
\begin{eqnarray}
S_{rb}=\frac{\omega_{max}-\omega_{min}}{N}\times10
\end{eqnarray}
in which $\omega_{max}$ and $\omega_{min}$ define the range of $\omega_i$ at the significance of 90\%
and $N$ is the number
of the data points within the range. The above equation means that on average 10 data points will
fall in the bin. We use it to guarantee that the generated histogram is of statistical significance.

The top panel on the right
column of Figure~\ref{fg_mc001} shows the histogram of $\omega_i$ by using this method under the best-fit
condition of the 1995 October MC. An outstanding peak is found in the histogram. We
locate the positions of the half maximum, $\omega_l$ and $\omega_r$ (as indicated by the two vertical
dashed lines). The optimized value of $\omega$ is then determined by $\frac{\sum\omega_iN_i}{\sum N_i}$
between the two half-maximum locations, and $\omega_l$ and $\omega_r$ give the uncertainties. For this case,
$\omega=4.85_{-1.67}^{+1.69}$. The next panel shows the parameter $\frac{B_\varphi}{B_z}$ as a function of
$x$ under the best-fit condition. The solid line corresponds to $\omega=4.85$ with the two dashed lines for
the uncertainties, and the data points within the uncertainties are highlighted in blue. How close the
magnetic field lines are to the uniform-twist configuration is then assessed by (1) the percentage of
the selected data points which locate within the uncertainties in all the data points in the
MC interval, (2) the correlation between the modeled and measured $\frac{B_\varphi}{B_z}$ of the selected data points
and (3) the confidence level of the correlation under the permutation test.
Again for this case, the percentage (per) is 56\% and the coefficient of correlation (cc) is 0.74 with the confidence level (cl)
of nearly 100\% as shown in the bottom panel on the right column of Figure~\ref{fg_mc001}.

\begin{figure*}[tb]
  \centering
  \includegraphics[width=\hsize]{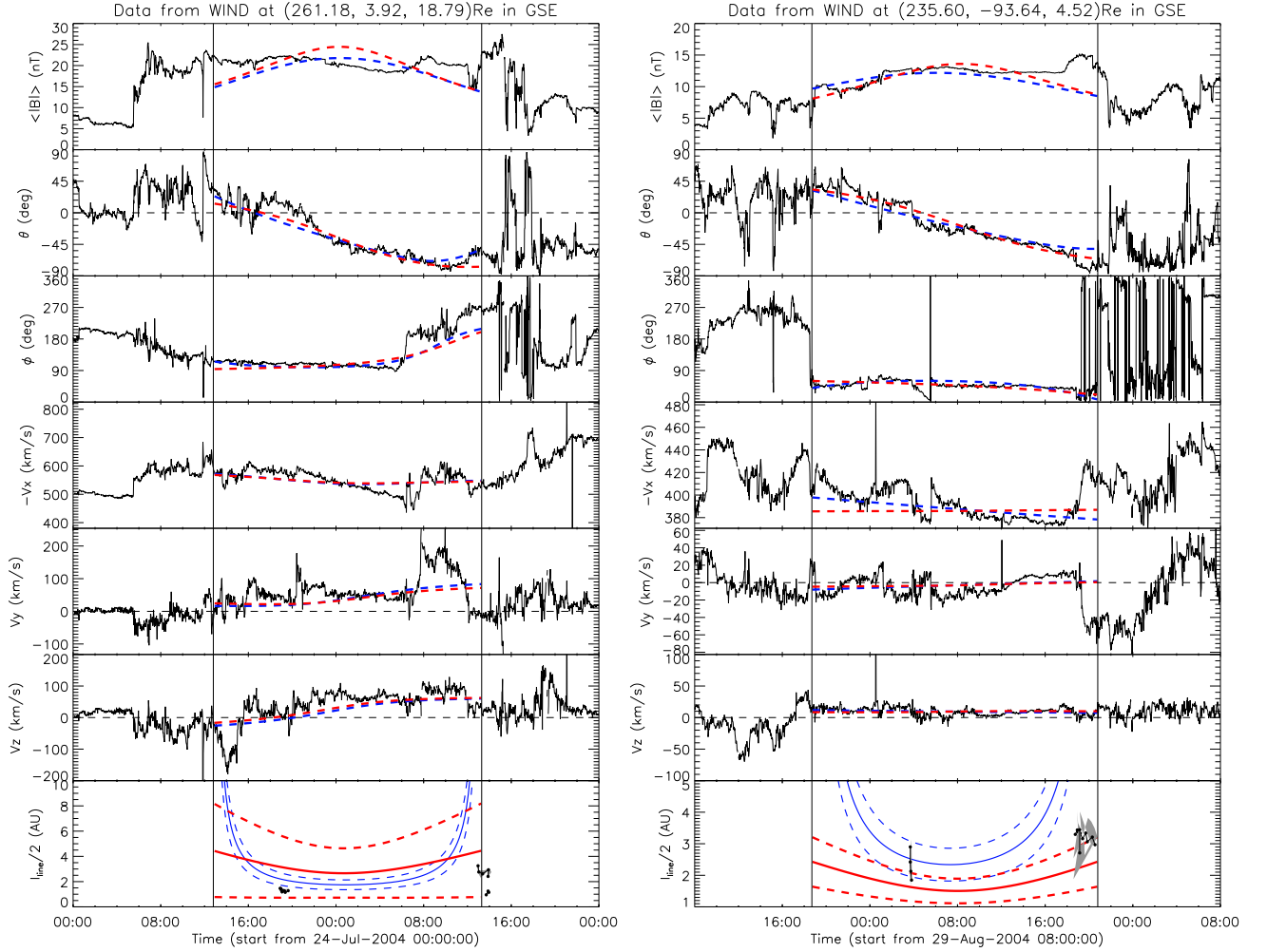}
  \caption{The Wind data and fitting curves for the MC Nos.6 and 7 with the same arrangement as the
left panel of Fig.\ref{fg_mc001}.}\label{fg_mc006-7}
\end{figure*}

Meanwhile, the goodness-of-fit is evaluated by
\begin{eqnarray}
\chi_n&=&\sqrt{\frac{1}{2N}\sum_{i=1}^N\left[\left(\frac{\mathbf{B}_i^m-\mathbf{B}_i^o}{|\mathbf{B}_i^o|}\right)^2+\left(\frac{\mathbf{v}_i^m-\mathbf{v}_i^o}{|\mathbf{v}_i^o|-v_{\rm ref}}\right)^2\right]} \nonumber\\
&=&\sqrt{\frac{1}{2}(\chi_{Bn}^2+\chi_{vn}^2)} \label{eq_chi}
\end{eqnarray}
where the superscript $m$ and $o$ denote the modeled and observed values, respectively, $N$ is the number of measurements,
and $v_{\rm ref}$ is a reference velocity. $\chi_n$ gives the overall relative error between the modeled and
the observed values. A much more detailed explanation of the equation of $\chi_n$ can be found in~\citet{Wang_etal_2015}.

Since the goodness-of-fit and the goodness of the uniform-twist configuration are assessed by two independent approaches,
they may not be positive correlated. We then use the
following two conditions to classify the quality, $Q$, of the model fit:
\begin{itemize}
\item $\mathrm{per}\geq 50\%$, $\mathrm{cc}\geq0.5$ and $\mathrm{cl}\geq 90\%$
\item $\chi_n\leq0.5$
\end{itemize}
If both conditions are satisfied, $Q$ is 1, and if only the second condition is satisfied, $Q$ is 2. As long as the second condition
is not satisfied, the fitting is treated to be completely failed.

\section{Comparison of the twists derived from the model with those from other methods}\label{sec_compare}
\subsection{Comparison with the electron probe method}
First, we will compare the magnetic field line lengths estimated by our model with those inferred
from the energetic electron probes. \citet{Kahler_etal_2011} studied 30 type-III burst-associated energetic
electron events, of which 16 events located within 8 MCs. We then focus on these MCs and fit them with
our model. It is found that all but one of the MCs can be fitted with three of them having $Q=1$ and four $Q=2$ (see
Table~\ref{tb_parbv}, event Nos.1--7). It is noted that the field line length, $L_e$, listed in Table 1 of
\citet{Kahler_etal_2011} is not the length from one end of the MFR to the other but that from the Sun to the
Wind spacecraft. Thus, based on our picture shown in Figure~\ref{fg_coord}, we use
$\frac{l_{line}}{2}$ with $\lambda=2.57\pm0.57$ and $L=1$ AU (see Eq.\ref{eq_length} and \ref{eq_l}) for the comparison
by assuming that the MCs were crossed at their apexes.

Three events, Nos.1, 6 and 7, with $Q=1$ are presented in Figure~\ref{fg_mc001} and \ref{fg_mc006-7}.
The red dashed lines superimposed on the magnetic field and velocity profiles are the fitting curves.
For comparison, the fitting curves of the velocity-modified Lundquist model~\citep{Wang_etal_2015} are
also plotted in blue.
The modeled magnetic field line lengths are given in the last panels by the solid lines with the
uncertainty in dashed lines. The uncertainties in the lengths for the velocity-modified GH model
come from two main sources; one from the
uncertainty in the axial length $l$ and the other from the uncertainty in $\omega$ or $\tau$ (see Eq.\ref{eq_length} or \ref{eq_dl}).
Those for the velocity-modified Lundquist model are estimated only from the uncertainty in the axial length.
For the 1995 October 18 MC, the lengths probed by energetic electrons are almost all within the
GH model range of the lengths, but close to the lower boundary. The change trend of the modeled
length with the time is consistent with that of the probed length. It is similar for the 2004 July 24 MC,
in which the probed lengths are close to the lower boundary of the GH modeled lengths.
For the 2004 August 29 MC, the probed lengths are slightly longer than the GH modeled lengths.

\begin{figure*}[tb]
  \centering
  \includegraphics[width=0.82\hsize]{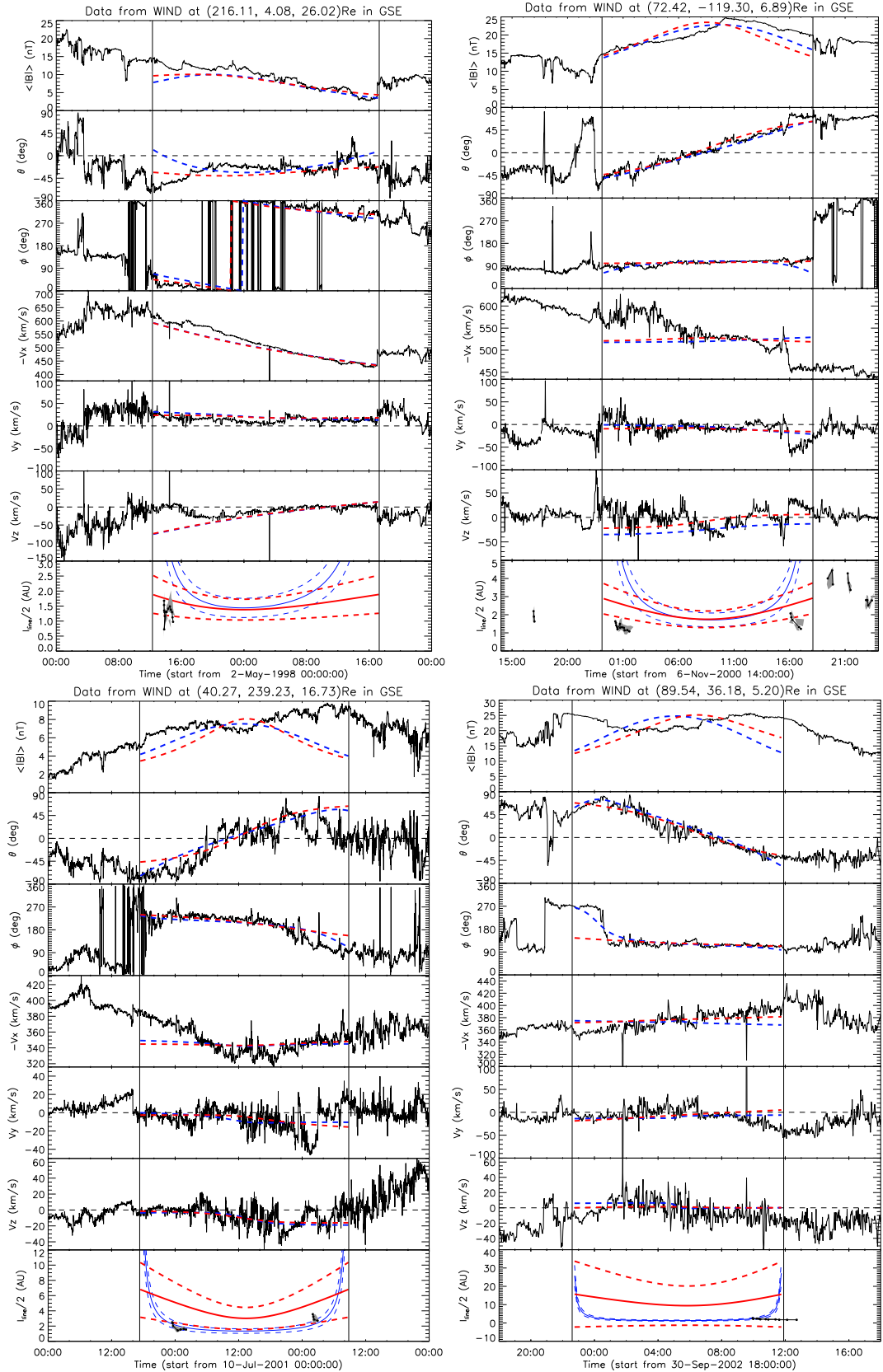}
  \caption{The Wind data and fitting curves for the MC Nos.2--5.}\label{fg_mc002-5}
\end{figure*}

The other four events with $Q=2$ are shown in Figure~\ref{fg_mc002-5}. The best match of the
GH modeled length with the probed length happens to the 1998 May 2 MC. The other three events
show more or less significant deviations between the modeled and probed lengths. For the
2000 November 6 and 2001 July 10 MCs, the probed lengths locate around or even outside of
the lower boundary of the GH modeled length. Together with the 2004 August 29 MC, these
MCs are not typical. Their radial velocities show a generally declining profile, which indicate
an expansion and therefore may result in a decreasing magnetic field with time, but the measured
total magnetic field strength somehow increased with time (for the 2001 July 10 MC, its first half
shows the inconsistency). For the last event on 2002 September 30, the uncertainty of the GH modeled
length is too large to be useful though the probed lengths fall in the modeled length range.
This MC is also non-typical. Its radial velocity was continuous increasing, but the total magnetic
field did show a declining profile.

Comparing to the fitting results of the velocity-modified Lundquist model (the blue lines in
Fig.\ref{fg_mc001}--\ref{fg_mc002-5}),
we find that the GH model is generally better than the Lundquist model, particularly near the periphery
of the MCs where the Lundquist model predicts extremely long field lines, well exceeding the probed lengths.
It confirms the conclusion of \citet{Hu_etal_2015} that the magnetic field lines in MCs are more likely to be uniformly twisted.
However, we want to note that the uncertainty of the velocity-modified GH model in
modeling the field line length sometimes is quite large, which is due to the two error sources mentioned above. Based on the
error propagation theory for absolute errors, we may estimate
\begin{eqnarray}
\Delta l_{line}&=&\frac{\partial l_{line}}{\partial l}\Delta l+\frac{\partial l_{line}}{\partial \tau}\Delta\tau\nonumber\\
&=&\sqrt{1+(Tr)^2}\Delta l+\frac{4\pi^2r^2l|\tau|}{\sqrt{1+(Tr)^2}}\Delta\tau \label{eq_dl}
\end{eqnarray}
according to Eq.\ref{eq_length}. As long as the observational path is not too close to the MFR axis, the variable $\sqrt{1+(Tr)^2}\approx|T|r$,
and the ratio of the first item on the right-hand side of Eq.\ref{eq_dl} to the second one is therefore approximately
$\frac{\Delta l/l}{\Delta\tau/|\tau|}$. Based on the discussion following Eq.\ref{eq_l}, we may infer that $\frac{\Delta l}{l}$
is about 0.22. Thus, for the MCs (Nos.1, 4--6) with $\frac{\Delta\tau}{|\tau|}$ larger than 0.22 (see column 16 of Table~\ref{tb_parbv}),
the uncertainty in the field line length mainly comes from the uncertainty
in the modeled twist. Particularly, the MC5 has an extremely large twist, which is unreliable (a stronger reason of its
unreliability can be seen in Fig.\ref{fg_tau_dis}a and Sec.\ref{sec_stat}).

\begin{figure*}[tb]
  \centering
  \includegraphics[width=\hsize]{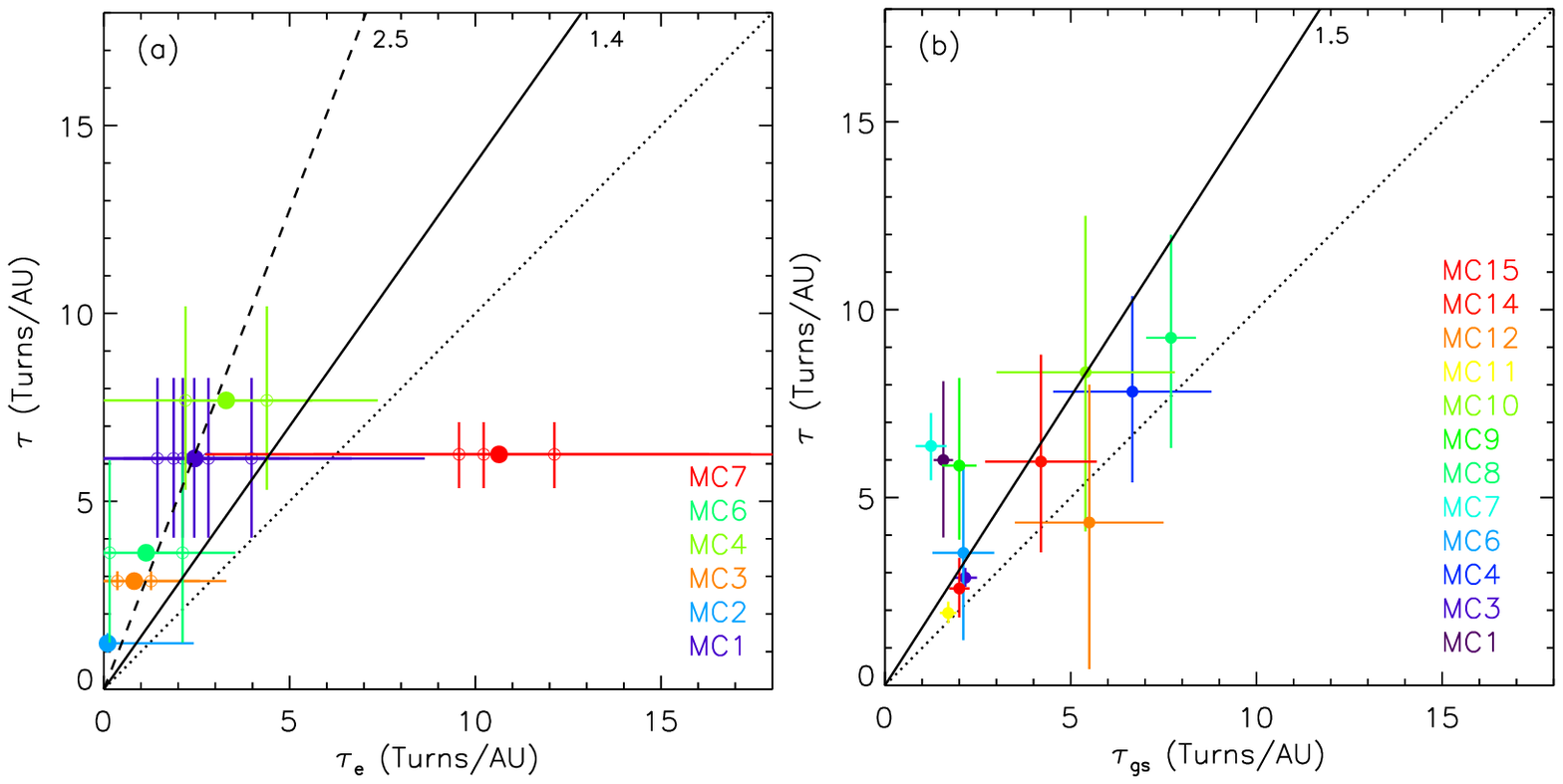}
  \caption{Comparison of the number of turns per AU, $\tau$, derived by the velocity-modified GH model with (a) that by
the electron probe method and (b) that by the GS model. Each circle in the left panel indicates an
energetic electron event and each solid dot marks the mean value of the twists of a MC. The solid lines are the linear
fits to the solid dots in both panels. The dashed line in the left panel is the linear fit to the solid dots excluding MC7.
The numbers on the top of the panels give the slopes of the linear fitting lines.}\label{fg_twist_comp}
\end{figure*}

\begin{figure*}[tb]
  \centering
  \includegraphics[width=\hsize]{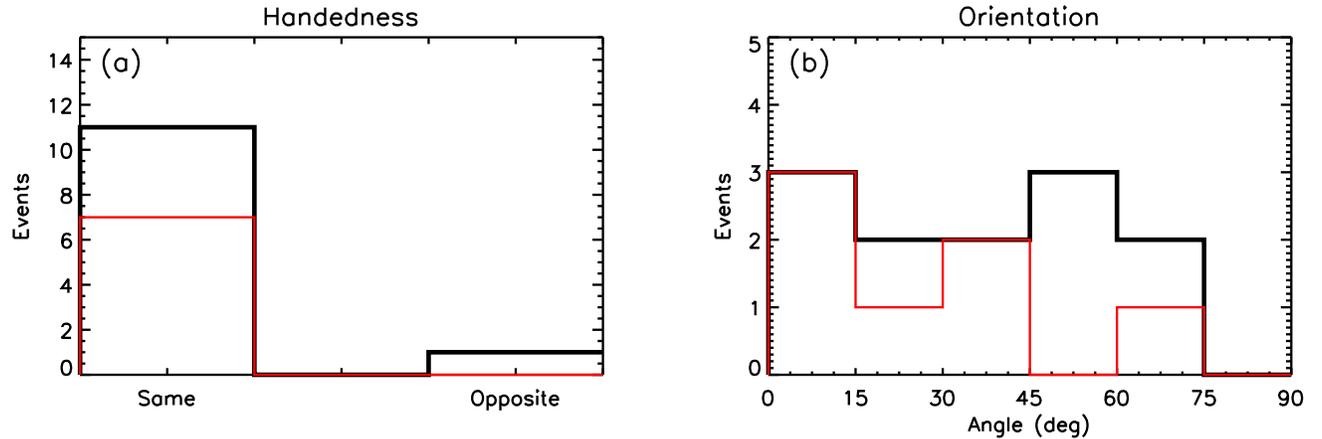}
  \caption{Comparison of the derived handedness and orientation between the velocity-modified GH model and GS model. The
black histograms are for the MC events Nos.1, 3, 4, 6--12, 14 and 15, and the red histograms for the MC events Nos.8--12,
14 and 15.}\label{fg_utvsgs}
\end{figure*}

On the other hand, we convert the probed magnetic field line lengths to the twists based on the
configuration of the GH flux rope by using Eq.\ref{eq_length} with the assumption that
the axial lengths of the MCs are $2.57\pm0.57$ AU, and compare these electron-probe-based
twists, $\tau_e$, with the GH modeled twists, $\tau$. Figure~\ref{fg_twist_comp}a
exhibits the result. Hereafter we use the absolute values of $\tau$ and $\tau_e$ in the figures.
Each circle indicates an energetic electron event, and each solid dot is the mean value
of the twists for a MC event.
Note that MC5 is not included in the figure. It is found that except for MC7, the modeled twists of all the other MCs are larger
than the electron-probe-based twists. Since the electron probe method uses less assumptions than
the GH fitting technique, we think that the electron-probe-based twist is closer to the real twist than the modeled twist.
By fitting the solid dots with the formula of $\tau=a\tau_e$ without
considering the uncertainties, we find that the modeled twist
is over-estimated by a factor of 1.4 on average (as indicated by the solid line) and the
correlation coefficient between the two sets of the twists is 0.59. If ignoring the dot for MC7, we find that all the other
dots almost align with each other. By using the same formula to fit the dots (as indicated by the dashed line),
we get that the over-estimation factor of
the modeled twist is about 2.5 and the correlation coefficient increases to 0.9987.
We tend to believe that the overall over-estimation factor of the modeled twist is more likely to be 2.5. It is
worthwhile to refine the number by searching more energetic electron events in the future.

\subsection{Comparison with the Grad-Shafranov reconstruction technique}\label{sec_gs}
By using GS reconstruction technique, Hu and his co-authors~\citep{Hu_etal_2014, Hu_etal_2015}
have modeled the twists of 25 MCs, which is combined from three sets of MC events. The first set
consists of 9 MCs from~\citet{Qiu_etal_2007}, the second set consists of 9 MCs from~\citet{Hu_etal_2014},
and the last set of 7 MCs is from~\citet{Kahler_etal_2011}. We do not include the first set of MCs
in the comparison here, because there are large disparity in determination of boundaries of the MCs among
literatures as mentioned in~\citet{Qiu_etal_2007}. Moreover, two of the rest of the MCs cannot be successfully
fitted by our model, and thus a total of 14 MCs are finally put in the comparison, which are the
MC Nos.1 and 3--15 listed in Table~\ref{tb_parbv} (Note that MC2 is not in the Hu's list).
It should be noted that MC5 and MC13 have an extremely large fitted
twist, which is unreliable.
We think that a direct reason causing such an extremely
large twist is the rather small radius of the MFR, which is only 0.01 AU. However, the GS reconstruction
gives much reasonable twists for the two MCs, which are $\tau_{gs}=4.2\pm0.54$~\citep[see Table 1 in][]{Hu_etal_2015} and
$14.6\pm5.4$~\citep[see Table 2 in][]{Hu_etal_2014}, respectively.

The comparison of the twists from the two models is shown in Figure~\ref{fg_twist_comp}b, in which
the data points of MC5 and MC13 are not included. The coefficient of the correlation between the two
sets of twist is about 0.68, with our GH twist larger than the GS twist by a factor of 1.5 on average, which
is smaller than the over-estimation factor of 2.5 found in the last section but very close to the over-estimation
factor of 1.4 obtained by including MC7. These two correlations
shown in Figure~\ref{fg_twist_comp} suggest that (1) on average the GS technique is more accurate than the velocity-modified GH
model to infer the twist of magnetic field lines of MCs though it still probably over-estimates the twist by a factor
of 1.7, and (2) the correlation between the GH modeled and probed twists is better than that between the
GS modeled and probed twists if MC7 was excluded. Moreover, it should be noted that for individual cases,
both GS and GH models might give a twist deviated largely from the probed twist.

To have a complete view about the similarity and difference of the fitting results between the GH and GS models,
we compare the other two key parameters: the handedness and orientation of the MCs. The
handedness, i.e., the sign of the helicity, is a fundamental parameter characterizing the topology
of a magnetic field system, and the orientation determines the configuration of the MC in 3D space,
and may significantly influence the geoeffectiveness of the MC~\citep{Wang_etal_2007a}.
The differences in the handedness and orientation between the two models have been shown in Figure~\ref{fg_utvsgs}.
The black histograms are for the common events Nos.1 and 3--15 except for the events Nos.5 and 13 which have unreasonably large
twists from the velocity-modified GH model. It is found that there is one event (the event No.4) getting
an opposite handedness from the two models, and the angles between the two modeled orientations almost uniformly scatter between
$0^\circ$ and $75^\circ$. It should be noted that for the events Nos.1--7, the boundaries of the MCs used in
\citet{Hu_etal_2015} are slightly different from those in Lepping's list. If we exclude those events, the
two models get better consistent results in the two parameters as shown by the red histograms. The handednesses
are all the same. In 6 of 7 events the difference between the two modeled orientations is less than $45^\circ$ and
more than half of the 7 events have a difference less than $30^\circ$. The comparison suggests that the
velocity-modified GH model is roughly consistent with the GS model, and the identification of the boundaries
of a MC is of importance to the fitting results.

\section{Statistical results}\label{sec_statistics}
Although the twist estimated by the velocity-modified GH model is probably 2.5 times of the real twist on average as revealed
in the above sections, we still can investigate the statistical properties of the twists of interplanetary MFRs
based on the model because of the high correlation ($cc\approx0.9987$) with the probed twists.
We apply the velocity-modified GH model to all of the 121 MCs in Lepping's list (see
\url{http://lepmfi.gsfc.nasa.gov/mfi/mag_cloud_S1.html}, including the 7 MCs from \citealt{Kahler_etal_2011}).
The MC Nos.45 and 46 in their list are not included because of the data gaps in the published Wind data, and
the MC No.85 is also removed because it is believed to consist of two MCs~\citep{Dasso_etal_2009}.
After setting the same boundaries of the MCs as those given in Lepping's list and the time resolution
to 10 minutes, the entire fitting procedure is automated. Plus the 8 MCs from \citet{Hu_etal_2015}, we have
a total of 126 MCs. We find that there are 115 ($\sim91\%$) of
these MCs with the fit quality $Q$
of 1 or 2, and 52 MCs (occupying $\sim41\%$) having $Q=1$,  which have been listed in Table~\ref{tb_parbv}.

\begin{figure*}[tb]
  \centering
  \includegraphics[width=0.8\hsize]{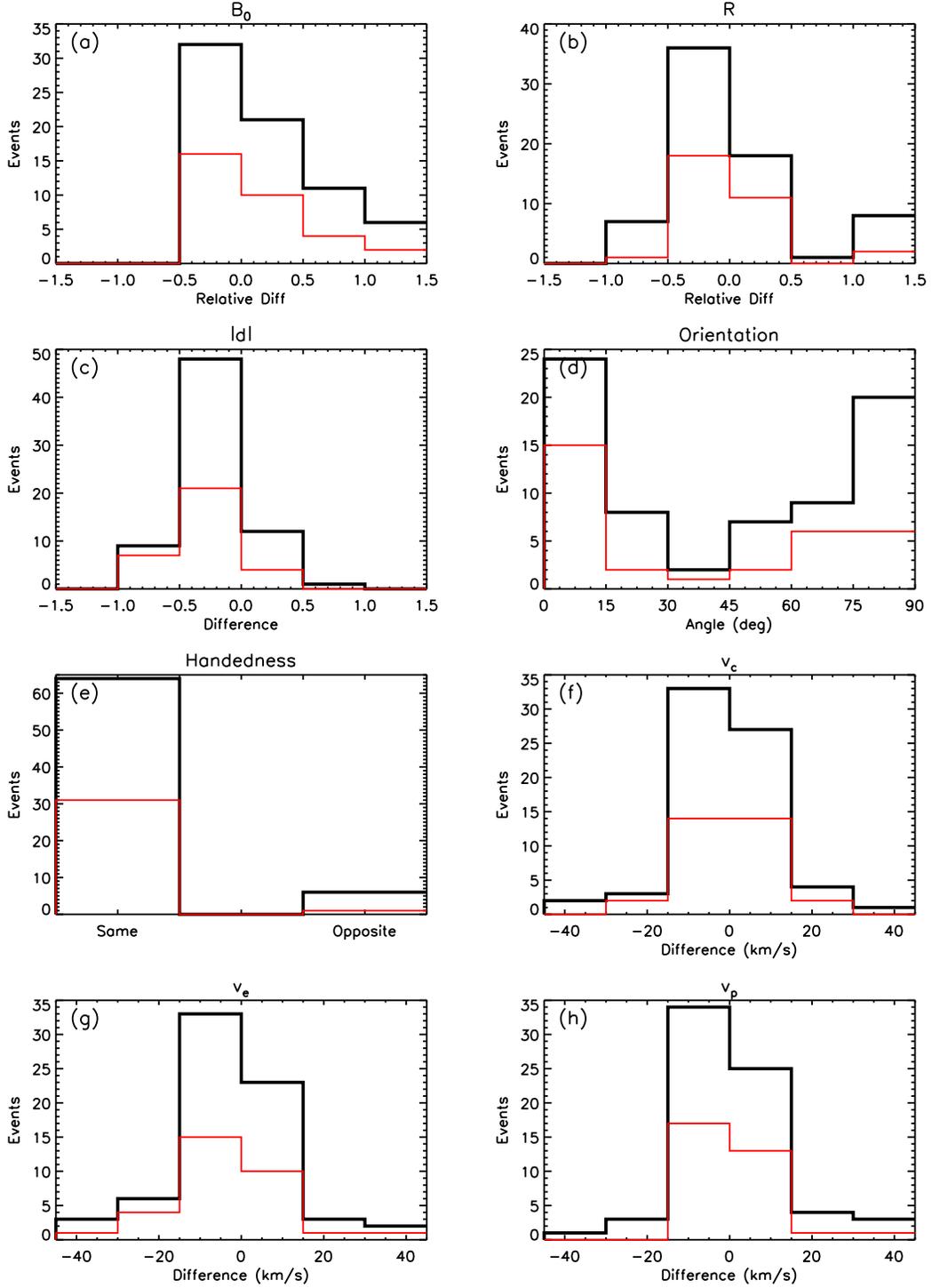}
  \caption{Histograms showing the differences of the values of fitting parameters between the velocity-modified GH model and
the velocity-modified Lundquist model. From left to the right and top to the bottom, they are (a) the relative
difference, i.e., $\frac{f_{gh}-f_{lq}}{f_{lq}}$, of the magnetic field strength ($B_0$) at the MCs axis between the two models, (b) the relative difference of the
radius (R), (c) the difference of the closest approach |d|, (d) the acute angle between the MCs's orientations,
(e) the difference of the handedness, (f) the difference of the propagation speed, (g) the difference of the expansion speed
and (h) the difference of the poloidal speed. The black histograms are for the events with both $Q_0$ and $Q$
equal to 1 or 2, and the red histograms for the events with $Q_0$ equal to 1 or 2 and $Q$ equal to 1.}\label{fg_utvslq}
\end{figure*}

\subsection{Comparison with the Lundquist model}\label{sec_lq}
Before analyzing the statistical properties of the twist, we compare the fitting results of the velocity-modified GH model with
those of the velocity-modified Lundquist model in \citet{Wang_etal_2015}. Figure~\ref{fg_utvslq} shows the
histograms of the differences in some parameters between the GH and Lundquist models. Since only the 72 MCs
with the quality $Q_0$ of 1 or 2 (listed in the last column of Table~\ref{tb_parbv}, an index used by~\citealt{Lepping_etal_2006}
to mark the fitting quality; $Q_0=1$ or 2 means good or fair) were studied in
\citet{Wang_etal_2015}, here we also only include those MCs with both $Q_0$ and $Q$ of 1 or 2 in the comparison,
which counts a total of 70 MCs. The black lines are drawn for all of these MCs and the red lines for those with $Q=1$.
First, the two sets of histograms look quite similar, suggesting that
the differences in the parameters between the two models do not depend on the quality of the fit. Second,
the fitting results of the two models are more or less different, and the difference may be significant for some parameters.
For the most of the MCs, the relative differences in
the total magnetic field strength, $B_0$, the radius, $R$, and the closest distance to the axis of the MC, $|d|$,
are between $\pm 0.5$ with the trend that the values derived from the GH model are slightly smaller than those from
the Lundquist model. There are a few cases showing an opposite handedness between the two models. The differences
in the modeled velocities are the most insignificant. The largest difference of the fitting results
appears in the orientation of the MC's axis. The black histogram reveals that the angle of between the two modeled
orientations is larger than $45^\circ$ for 36 of the 70 MCs. Even if only best-fitted MCs were considered,
there are 14 of the 32 MCs do not match well in the orientation (see the red histogram). But we still can
find that the two models got quite consistent orientations for $1/3$ to $1/2$ of the MCs.

Orientation is one of the most important parameter of interplanetary MFRs. \citet{Riley_etal_2004} performed
`blind tests' by applying five different fitting techniques to a MHD simulated MC. It was found that the deviation
in the orientation among these different models is quite significant, especially when the observational path
is far away from the axis of a MC, i.e., $|d|$ close to unity. There is so far no direct or indirect observations
to justify which model is better for a given MC. However, the tests by \citet{Riley_etal_2004} did show that the fitting technique based
on a force-free flux rope is a useful tool. Moreover, the comparison between the velocity-modified GH model and
the GS model in Sec.\ref{sec_gs} already implies that our model results are acceptable.
Besides, in our sample (see Table~\ref{tb_parbv}), there are 4 events
with $|d|>0.9$ or 8 events with $|d|>0.8$, two of which have a twist larger than 20 turns per AU and will be excluded in the statistical
analysis below. Thus even if the modeled parameters of the event with large $|d|$ are much unreliable, the influence of
the few events on the statistical result would be small.

\subsection{Statistical properties of the twist}\label{sec_stat}
\begin{figure*}[tb]
  \centering
  \includegraphics[width=\hsize]{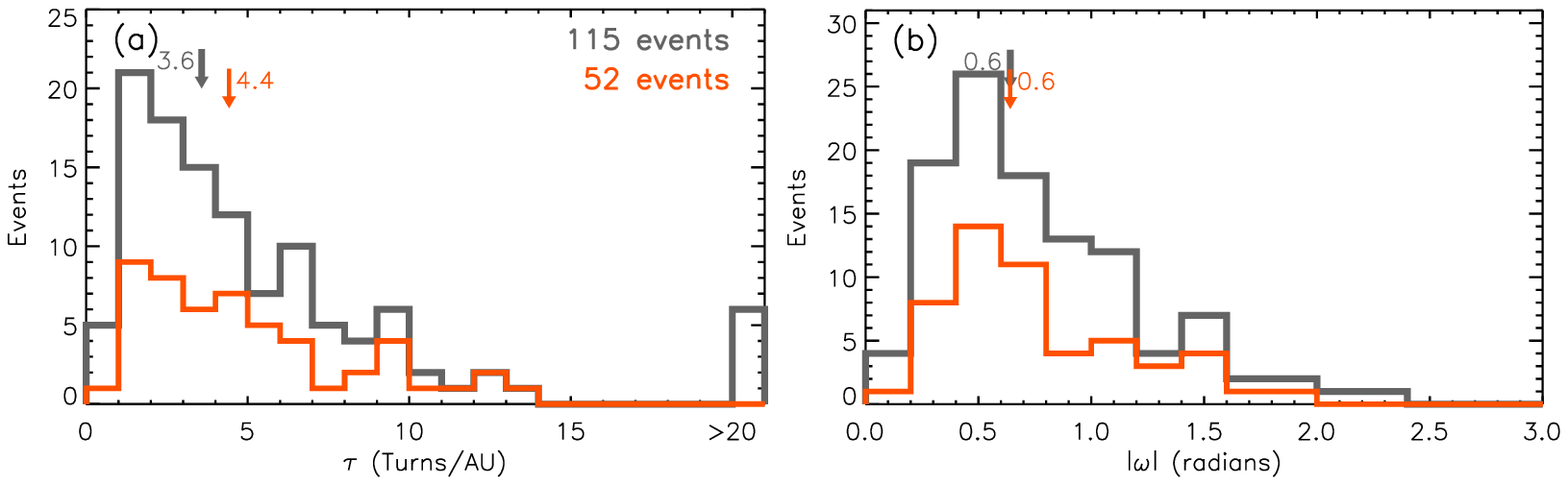}
  \caption{Distributions of the derived (a) $\tau$ and (b) $\omega$ for all of the events
(in gray color) and for the events with $Q=1$ (in orange color). The arrows mark the median values. The events with
$\tau > 15$ turns per AU are not counted in Panel (b). The over-estimation effect is corrected in only Panel (b).}\label{fg_tau_dis}
\end{figure*}

Figure~\ref{fg_tau_dis}a shows the distribution of the twists of the 115 MCs listed in Table~\ref{tb_parbv}
in terms of $\tau$. The peak of the distribution locates between $\tau$ of
1 and 2. From the peak, the number of events decreases with increasing $\tau$,
and reaches zero before $\tau=15$ turns per AU. There are six events with $\tau>20$ turns per AU, which look clearly not following the trend
of the main part of the distribution, suggesting that these events are not successfully fitted by the
velocity-modified GH model though their qualities assessed based on the criteria listed in Sec.\ref{sec_quality}
are all equal to 2. This gives the reason why we excluded event Nos.5 and 13 in Figure~\ref{fg_twist_comp}.
For the rest of the events, the median value of $\tau$ is about 3.6. Besides, there are only five events with
$\tau<1$ turn per AU, occupying 4\% of the events.

If only considering the events with $Q=1$, which forms a sample of 52 events, we find that the distribution
is similar, as shown by the orange line in Figure~\ref{fg_tau_dis}a. The most probable value of $\tau$ is between
1 and 2 turns per AU, and the median value of $\tau$ slightly increases to 4.4 turns per AU. Only one event
locates in the bin of $\tau<1$ turn per AU.
Assuming that interplanetary MFRs still attach both ends to the Sun, which implies the shortest
axial length of 2 AU from its one end to the other, and considering that the over-estimation factor of $\tau$ is about
2.5, we may infer that the most probable value of the total twist angle, $\Phi_T$, of a MFR is between $1.6\pi$ and $3.2\pi$, and the
median value is about $5.7\pi-7.0\pi$. It implies that (1) a significant fraction ($>80\%$,
read from Fig.\ref{fg_tau_dis}a) of
the MFRs possess highly twisted magnetic field lines exceeding the HP critical twist, which is $2.5\pi$,
for the kink instability of a line-tying GH flux rope, and (2) a few MFRs almost did not carry twisted
magnetic field lines.

It was mentioned in Sec.\ref{sec_model} that \citet{Hood_Priest_1981} derived the HP critical twist
based on the GH flux rope with the assumption of $\omega=1$, i.e., the radius $R=\frac{1}{T}$. Thus, it is interesting to see how well
the assumption matches the observation-based model results. Figure~\ref{fg_tau_dis}b shows the distribution
of the absolute value of $\omega$ for the MCs (the over-estimation effect has been corrected by assuming the same over-estimation
factor of 2.5). The distribution has a peak around $\omega=0.5$ with a decrease
toward the larger-$\omega$ side and the median value is about 0.6. It could be estimated that
only about 20\% of the MCs have a modeled $\omega$ within the range of $1.0\pm0.2$.
Particularly, the value of 2 (0.2) seems to be an upper (a lower) limit of $\omega$ (see the next paragraph) .
Obviously, $\omega=1$ is not a good assumption for the most of the MCs.

\begin{figure*}[tb]
  \centering
  \includegraphics[width=\hsize]{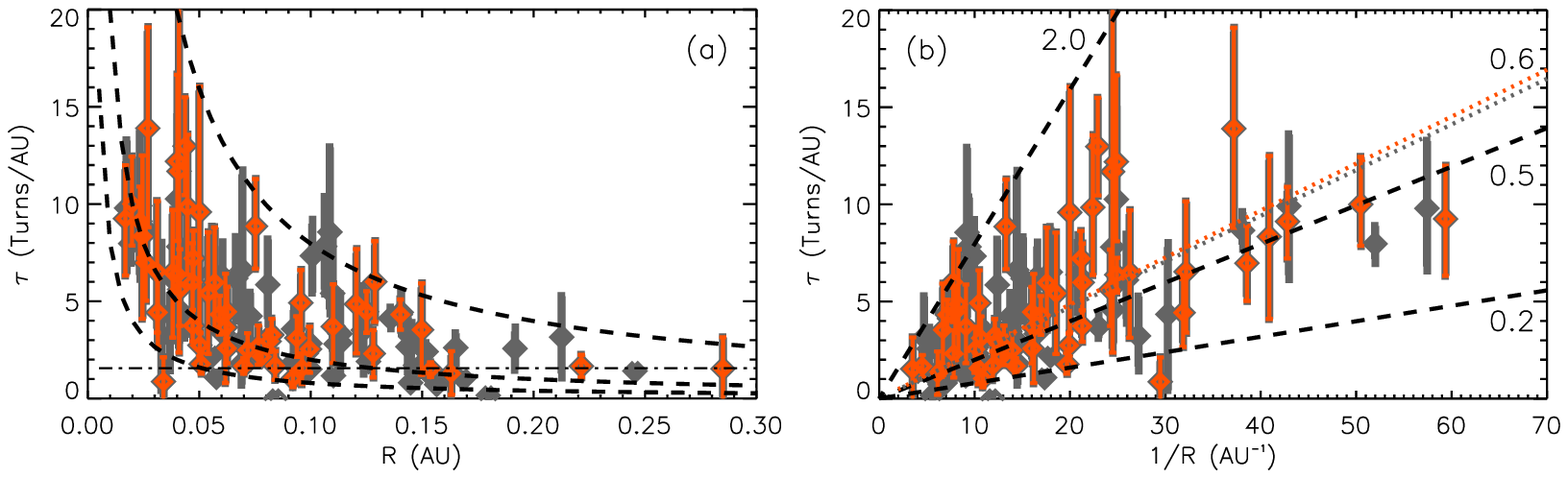}
  \caption{Scatter plots showing $\tau$ as a function of $R$ and $1/R$, respectively. The events with
$\tau > 15$ turns per AU are not included. The colors have the same meaning as that in Fig.\ref{fg_tau_dis}.
The dashed lines in both panels give $\tau=\frac{\omega}{2\pi R}$, equivalent to $\Phi_T=\omega\frac{l}{R}$, with $\omega=2.0$, 0.5 and 0.2.
Note, the over-estimation factor of 2.5 in the modeled $\tau$ has been considered in plotting these lines.
The dot-dashed line in panel (a) marks the HP critical twist by assuming the shortest axial length of 2 AU (a
correction of the factor of 2.5 is also considered), and the dotted lines in panel (b) are the linear fits to the data points}\label{fg_tau-r}
\end{figure*}

According to Eq.\ref{eq_tau} and \ref{eq_r-omega}, we may get $\tau=\frac{\omega}{2\pi R}$ or $\Phi_T=\omega\frac{l}{R}$, which
has the same form of Eq.\ref{eq_critical}. The above formula implies that
$\tau$ is perhaps proportional to $1/R$, as the value of $\omega$ is unimodal distributed with a relatively narrow
width. Figure~\ref{fg_tau-r} shows the relations between $\tau$ and $R$ and between $\tau$ and $1/R$ for all the events with $\tau<15$ turns
per AU (the high-quality events of $Q=1$ are in orange). The patterns
of the total events look similar to those of the high-quality events.
The correlation between $\tau$ and $R$ is clear; a thinner MC tends to have
more turns of magnetic field lines. The linear fitting to the ($\tau$, $1/R$) data points suggests a
slope of 0.6 (the dotted lines in Fig.\ref{fg_tau-r}b), which is the same as the median value of $\omega$.
The correlation coefficient is above 0.62. Besides,
according to the $\omega$ distribution given in Figure~\ref{fg_tau_dis}b,
we plot three dashed lines for the characteristic $\omega$ values of 2.0, 0.5 and 0.2, respectively, in Figure~\ref{fg_tau-r}
(the over-estimation factor of 2.5 has been taken into account).
The three lines do demonstrate the upper and lower boundaries and the spine of these data points.
It is noteworthy that the upper limit, $\tau_c=\frac{1}{\pi R}$, is the same as the theoretical results
of \citet{Dungey_Loughhead_1954} and \citet{Bennett_etal_1999}, which predicted that the total
critical twist angle is two times of the aspect ratio of the MFR, i.e., Eq.\ref{eq_critical} with $\omega_c=2$.
No MC exceeding $\tau_c$ suggests that Eq.\ref{eq_critical} with $\omega_c=2$ is probably a sufficient condition
for the unstableness of MFRs, i.e., a MFR becomes absolutely unstable when $\tau_c$ or $\Phi_c$ is satisfied.
In contrast, most of the modeled twists exceed the HP critical twist (the dot-dashed line in Fig.\ref{fg_tau-r}a)
even if we only chose the MCs with $\omega$ around the unity. It might suggest that the HP critical twist
is more likely to be a special condition for the kink instability, which is only applied for a certain
configuration of MFRs. More discussions of this result will be given in Sec.\ref{sec_discussion}.

\begin{figure*}[tb]
  \centering
  \includegraphics[width=\hsize]{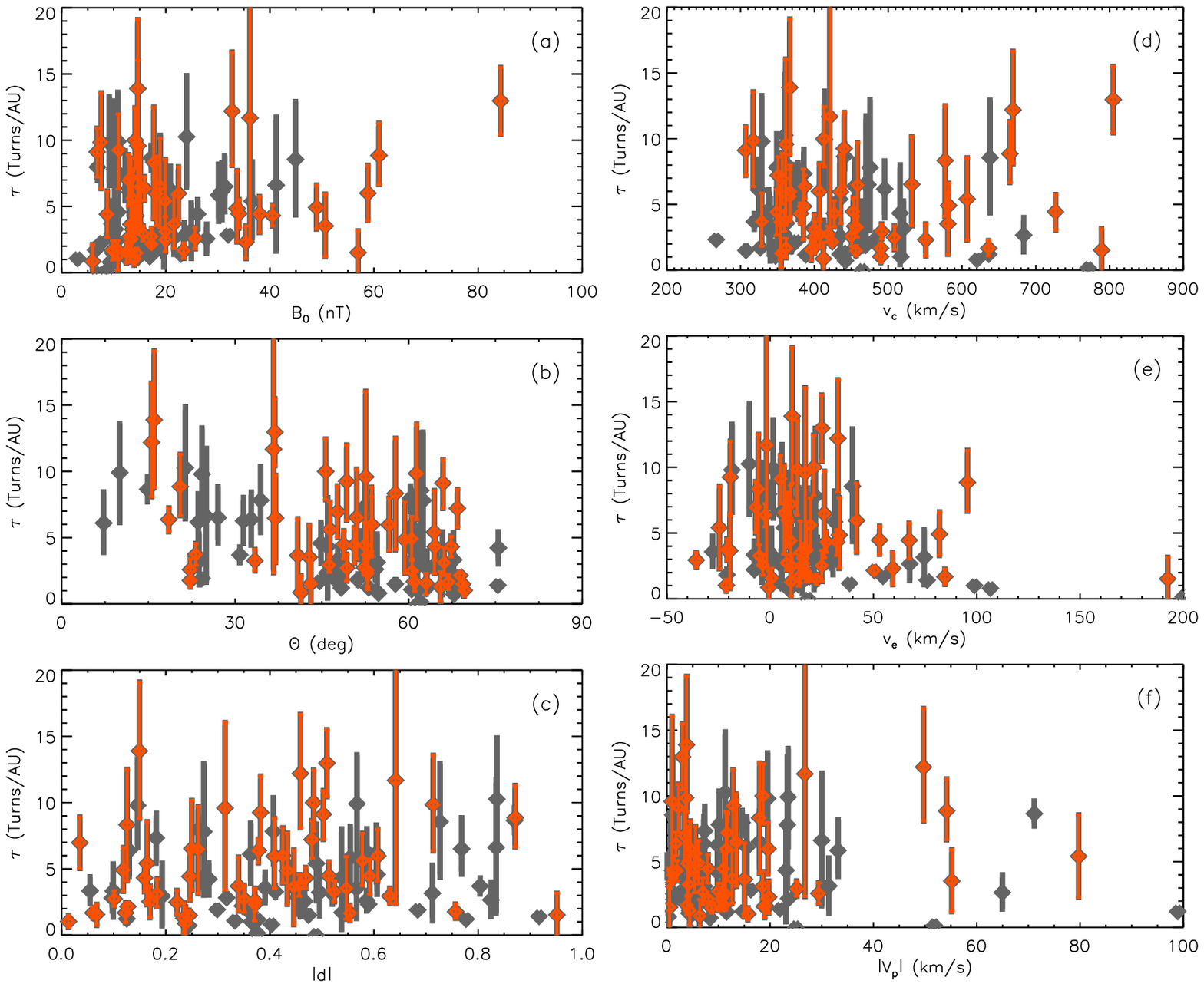}
  \caption{Similar to Fig.\ref{fg_tau-r}, but for (a) the magnetic field strength, $B_0$,
at the MC's axis, (b) the orientation, $\Theta$, of the MC's axis, (c) the closest
approach, $|d|$, between the observational path and the MC's axis, (d) the propagation speed, $v_c$,
(e) the expansion speed, $v_e$, and (f) the poloidal speed, $|v_p|$, of the MC.}\label{fg_tau_corr}
\end{figure*}

The relations between $\tau$ and other characteristic parameters of the MCs are shown in Figure~\ref{fg_tau_corr}.
There are no obvious correlations with $\tau$ except for the expansion speed and axial orientation.
One can find that the values of $\tau$
of the MCs with the larger expansion speed are not too large (Fig.\ref{fg_tau_corr}e).
Concretely speaking, there is no MC with $\tau>10$
turns per AU or there is only one MC with $\tau>5$ turns per AU among 15 MCs whose expansion speeds are
larger than 50 km s$^{-1}$. This phenomenon could be interpreted as that the twisted magnetic field lines constrain the expansion
of a MFR, then the size of a MFR; a MFR possessing a strong twist cannot be too thick. This does provide an alternative interpretation for
the dependence of the twist on the radius presented before in Figure~\ref{fg_tau-r}.
Although the result here suggests that the twist has effect on the size and expansion of interplanetary MFRs, it should be noted
that the ambient solar wind might play an even more important role in controlling their size and
expansion~\citep[e.g.,][]{Demoulin_Dasso_2009, Gulisano_etal_2010}. For those MCs with significantly
negative expansion speeds, \citet{Wang_etal_2015} have
shown that they were compressed by following fast solar wind streams.

The other possible correlation is between $\tau$ and $\Theta$ (Fig.\ref{fg_tau_corr}b). The parameter
$\Theta$ is the angle between the MC's axis and the Sun-Earth line. It is roughly revealed that the larger $\Theta$
is the smaller is the value of $\tau$. Based on the picture that a MC is a loop-like structure with two ends rooted
on the Sun, we may infer that $\Theta=0^\circ$ means that the leg of the MC is crossed by the spacecraft and
$\Theta=90^\circ$ that the leading part of the MC is crossed. Thus, the above result implies that the magnetic field
lines in the legs are more twisted than those near the apex. \citet{Demoulin_etal_2016} got a similar
result based on the Lundquist model. They argued that it could be a result of an observational bias. However,
it is still possible that such a non-uniform distribution of the twist along the MFR's axis is real. Observations
and modeling of solar MFRs may provide useful information to solve this puzzle, which is worth to be done in the future.

\section{Summary and conclusions}

In this study, we established a velocity-modified GH model.
By applying this model to previously studied MC events and comparing the modeled results with
those by the electron probe method and the GS reconstruction technique, we have shown that the model
can provide useful information of the length and twist of magnetic field lines in MCs, but
the modeled values of them are probably over-estimated by a factor of about 2.5.
We also showed that the modeled results of the velocity-modified GH model are comparable to those by
the GS technique and the cylindrical flux rope model with the Lundquist solution
though large differences can be found in some parameters for some cases.

Further, by applying the velocity-modified GH model to 115 MCs, consisting of the MC events in
Lepping's list and those studied in \citet{Hu_etal_2015}, we investigated the statistical
properties of the twists of MCs. The following interesting results are found:
\begin{enumerate}
\item Based on the criteria used in this work, about 91\% of MCs can be roughly fitted by the velocity-modified GH model,
among which half events can be fitted fairly well. The fitting results are close to the GS model results.

\item The distribution of the twist, $\tau$, i.e., the number of turns per AU, decreases from its peak
locating between $\tau$ of 1 and 2 to zero before $\tau$ reaching 15 with a median value of about 5 (Fig.\ref{fg_tau_dis}a).
This distribution reveals that all of the interplanetary MFRs possess a twist $T<12\pi$ rad AU$^{-1}$
or $\tau<6$ turns per AU with the over-estimation factor of 2.5 taken into account.

\item Most ($>80\%$) of the MCs have a total twist angle larger than the HP critical twist, $2.5\pi$.
The modeled twists generally follow the function $\tau=\frac{0.6}{2\pi R}$, equivalent to $\Phi_T=0.6\frac{l}{R}$, and
are bounded by $\Phi_T=2\frac{l}{R}$ and $0.2\frac{l}{R}$, which apparently
define the upper and lower limits of the twists (Fig.\ref{fg_tau-r}). These results
suggest that (1) Eq.\ref{eq_critical} with $\omega_c=2$ gives the sufficient condition for
the unstableness of MFRs, above which MFRs becomes absolutely unstable, (2) thinner MFRs have a higher instability threshold than
thicker MFRs, and (3) most CME flux ropes probably erupt before the sufficient condition is satisfied.

\item The MFRs with large expansion speeds are unlikely to have a large twist (Fig.\ref{fg_tau_corr}e).
Together with the dependence between $\tau$ and $R$,
it is implied that strongly twisted magnetic field lines probably limit the expansion and size of a MFR.

\item A weak correlation (Fig.\ref{fg_tau_corr}b) is found between $\tau$ and the angle between the
MFR's axis and the Sun-Earth line, $\Theta$. Roughly,
the larger $\Theta$ is, the smaller is the value of $\tau$, implying that the magnetic field lines in the legs wind around
the main axis more tightly than those in the leading parts of MFRs.
\end{enumerate}

\section{Discussion}\label{sec_discussion}
Interplanetary MCs come from the Sun and belong to post-eruption MFRs. The twist information derived from the interplanetary
MFRs more or less reflects the twist of solar MFRs, which are in pre-eruption stage.
In the light of the findings listed above, two points raised in Introduction are further discussed
below.
\begin{enumerate}
\item {\it The critical twist for unstableness.} The most interesting finding of our study is that the modeled twists of MCs significantly
exceed the HP critical value and are apparently bounded by Eq.\ref{eq_critical} with $\omega_c=2$.
First, the modeled twists of MCs are based on GH fitting technique,
which assumes that the MFR is in force-free state, whereas Eq.\ref{eq_critical} is derived for a non-force-free
flux rope. Why do the twists from the
two different methods and configurations have such an apparently connection? We do not have the answer at present
but it deserves a further study in future work.

Second, why is the HP critical twist much smaller than the modeled twists of MCs? In theory, the HP critical twist
was derived based on the GH flux rope with the assumption $\omega=1$. However, about 80\% of the MCs
have the value of $\omega$ other than 1, i.e., less than 0.8 or larger than 1.2 (see Fig.\ref{fg_tau_dis}b).
Moreover, considering $\Phi_T=\omega\frac{l}{R}$, the HP
critical twist does imply that a GH flux rope becomes unstable when the aspect ratio exceeds $2.5\pi$, which seems
to be away from the observed solar MFRs. Thus, these inconsistencies suggest that
the assumption is not good enough. However, as mentioned before, even if we only considered the MCs with $\omega\approx 1$, there
are still many MCs with the
modeled twists larger than the HP critical twist. This probably implies that most MCs are not exactly in the GH
configuration though the GH model can recover some useful information. Since all the theoretical analyses of the
instability were for MFRs in a stable initial state, the above discussion is valid only when
the modeled twist of the MCs are roughly the same as that before the MCs erupted from the Sun.

Actually, it was argued before that a significant fraction of the magnetic flux of a MC can be resulted from the magnetic
reconnection beneath the MFR during the eruption~\citep[e.g.,][]{Qiu_etal_2007}. This process may convert ambient
overlying fields either to the both poloidal and toroidal fluxes of the erupting MFR, adding a small or insignificant twist into
the MFR after the eruption~\citep{vanBallegooijen_Martens_1989}, or mainly to the poloidal flux, adding a large
twist~\citep{Longcope_Beveridge_2007, Qiu_2009, Aulanier_etal_2012}. Thus, it is possible that the eruption of a MFR
is firstly triggered by the kink instability at the HP critical twist~\citep[as suggested by, e.g.,][]{Fan_2005, Kliem_etal_2010},
and then the following reconnection process increases the twist to a much high level. If
the high twists found in the most MCs in this study were indeed mainly formed during the eruption, the newly-formed twist seems
obey Eq.\ref{eq_critical}, implying that the reconnection process will be interrupted when the total twist angle reaches $2\frac{l}{R}$
if it had not stopped earlier.

On the other hand, erosion process may occur to MCs during their propagation~\citep[e.g.,][]{Dasso_etal_2006, Ruffenach_etal_2012,
Manchester_etal_2014}, which progressively peels off the periphery of MCs from the front or rear
and is believed to cause on average about 40\% imbalance in the poloidal magnetic flux between the first and second half of a MC~\citep{Ruffenach_etal_2015}. This effect will make the modeled total flux underestimated, but might do little to the twist
for a uniform-twist flux rope. However, in practice, the erosion will more or less affect the values of fitting parameters including
the twist, but it is unclear how significantly the modeled twist is affected. The reconnection process during the
eruptions of MFRs and the erosion process during their propagations make the connection between the interplanetary MFRs and
the solar MFRs much loose.

\item {\it The inconsistency between the inferred twists of solar MFRs from imaging observations and those from modeling.}
Apparently, the modeled results of interplanetary MFRs are close to the twists estimated from the imaging data
but larger than those by using NLFFF extrapolation. As pointed out in Sec.\ref{sec_smfr}
the twist calculated based on the force-free parameter $\alpha$ is under-estimated for the uniform-twist flux ropes. Since the twist
of solar MFRs is more or less uniformly distributed as revealed by many studies~\citep[e.g.,][]{Inoue_etal_2011, Inoue_etal_2012,
Guo_etal_2013, Chintzoglou_etal_2015, LiuR_etal_2016}, it might be true that in these studies, the inferred twist is
significantly under-estimated. But it is not clear if the NLFFF extrapolation techniques applied in these studies
cause any under-estimation effects.
However, on the other hand, if most part of twist were resulted from the magnetic field
reconnection during eruptions as discussed in the previous point, the twist inferred from modeling, e.g., NLFFF extrapolations,
may be reasonable before and after but not during the eruptions, and the twisted structures
observed in multi-wavelength images might not really reflect the twisted magnetic structures.

\end{enumerate}


\begin{acknowledgments}
We acknowledge the use of the data from Wind spacecraft. We thank Stephen Kahler from AFRL, USA for
providing the data about the magnetic field line lengths inferred from energetic electron events. We also thank
the anonymous referees for useful comments. The model
developed in this work can be run and tested online at \url{http://space.ustc.edu.cn/dreams/mc_fitting/}.
This work is supported by the grants from NSFC (41131065, 41574165, 41421063, 41274173, 41474151)
CAS (Key Research Program KZZD-EW-01-4), MOEC (20113402110001)
and the fundamental research funds for the central universities.
\end{acknowledgments}

\appendix
\section{List of some acronyms}

\begin{itemize}
\item CME --- Coronal Mass Ejection
\item GH --- \citet{Gold_Hoyle_1960}
\item GS --- Grad-Shafranov \citep{Hu_Sonnerup_2002}
\item HP --- \citet{Hood_Priest_1981}
\item MC --- Magnetic Cloud
\item MFR --- Magnetic Flux Rope
\item NLFFF --- Non-Linear Force-Free Field
\end{itemize}

\section{List of the MC events}\label{app_tables}
\begin{table*}[h]\centering
\linespread{1} \caption{Parameters of the MCs involved in the study}
\label{tb_parbv}
\begin{tabular}{c|cc|cccccccccccccc|c}
\hline
 & \multicolumn{2}{c|}{MC Interval} & \multicolumn{14}{c|}{Modeled Parameters} &  \\
\cline{2-3}\cline{4-17}
No. & $t_0$ & $\Delta t$ & $B_0$ & $R$ & $\theta$ & $\phi$ & $d$ & $v_X$ & $v_Y$ & $v_Z$ & $v_{e}$ & $v_{p}$ & $\Delta t_c$ & $\Theta$ & $\tau$ & $Q$ & $Q_0$ \\
(1) &
(2) &
(3) &
(4) &
(5) &
(6) &
(7) &
(8) &
(9) &
(10) &
(11) &
(12) &
(13) &
(14) &
(15) &
(16) &
(17) &
(18) \\
\hline
\multicolumn{18}{c}{Events from \citet{Kahler_etal_2011}}\\
\hline
  1 & 1995/10/18 19:48 & 29.5 & 59 & 0.13 & -8 & 233 & 0.42 & -406 & -3 & 16 & 8 & 4 & 14.4 & 53 & 6.0$_{-2.1}^{+2.1}$ & 1 & 1 \\
  2 & 1998/05/02 12:18 & 29.0 & 10 & 0.17 & -60 & 344 & 0.36 & -516 & -15 & -26 & 98 & 4 & 11.6 & 61 & 1.0$_{-0.2}^{+0.2}$ & 2 & 3 \\
  3 & 2000/11/06 23:06 & 19.0 & 32 & 0.11  & 8 & 60 & 0.45 & -513 & -16 & -9 & 1 & -16 & 9.5 & 60 & -2.9$_{-0.3}^{+0.2}$ & 2 & 2 \\
  4 & 2001/07/10 17:18 & 39.5 & 19 & 0.11  & -15 & 148 & 0.41 & -348 & -11 & -5 & -3 & 10 & 20.1 & 34 & -7.8$_{-2.5}^{+2.4}$ & 2 & 2 \\
  5 & 2002/09/30 22:36 & 13.3 & 183 & 0.03  & -7 & 202 & 0.59 & -375 & -6 & -8 & -15 & -4 & 7.2 & 23 & -69.9$_{-64.6}^{+65.1}$ & 2 & 3 \\
  6 & 2004/07/24 12:48 & 24.5 & 51 & 0.15  & -29 & 32 & -0.55 & -576 & 72 & 16 & 14 & -55 & 11.9 & 42 & 3.5$_{-2.3}^{+2.4}$ & 1 & 2  \\
  7 & 2004/08/29 18:42 & 26.1 & 16 & 0.04  & -7 & 16 & -0.38 & -386 & -1 & 8 & -2 & -1 & 13.4 & 18 & 6.4$_{-0.9}^{+0.9}$ & 1 & 1  \\
\hline
\multicolumn{18}{c}{Events from \citet{Hu_etal_2015}}\\
\hline
  8 & 2008/03/08 19:20 & 5.5 & 11 & 0.02  & 11 & 145 & -0.38 & -437 & 45 & -25 & -19 & 12 & 2.9 & 36 & 9.3$_{-2.9}^{+2.7}$ & 1 &   \\
  9 & 2010/05/28 19:50 & 19.2 & 30 & 0.08  & -30 & 325 & 0.55 & -375 & -15 & -40 & 31 & 33 & 8.8 & 44 & -5.9$_{-2.3}^{+2.0}$ & 2 &   \\
 10 & 2010/08/04 04:00 & 4.0 & 18 & 0.02 & 33 & 230 & 0.13 & -576 & 30 & 2 & -5 & 17 & 2.1 & 58 & 8.3$_{-4.2}^{+4.2}$ & 1 &   \\
 11 & 2011/03/30 00:35 & 30.8 & 14 & 0.09 & 29 & 147 & 0.30 & -349 & 0 & -5 & 18 & 3 & 14.2 & 42 & 1.9$_{-0.3}^{+0.3}$ & 2 &   \\
 12 & 2011/06/05 01:20 & 5.1 & 20 & 0.03  & 53 & 318 & -0.54 & -515 & 7 & -13 & 15 & 23 & 2.4 & 63 & 4.3$_{-3.9}^{+3.7}$ & 2 &   \\
 13 & 2011/08/05 20:10 & 2.0 & 38 & 0.01  & -17 & 3 & -0.35 & -603 & 20 & -69 & 28 & -27 & 0.9 & 17 & 43.5$_{-3.0}^{+2.9}$ & 2 &   \\
 14 & 2011/09/17 15:45 & 14.1 & 18 & 0.06  & 30 & 40 & 0.41 & -434 & -20 & -18 & 41 & -11 & 6.2 & 49 & -6.0$_{-2.8}^{+2.4}$ & 1 &   \\
 15 & 2011/10/25 00:30 & 12.0 & 26 & 0.06  & 36 & 119 & -0.37 & -447 & 45 & -4 & -3 & 29 & 6.1 & 66 & -2.6$_{-0.8}^{+0.8}$ & 1 &   \\
\hline
\multicolumn{18}{c}{Events from \citet{Lepping_etal_2006}}\\
\hline
 16 & 1995/02/08 05:48 & 19.0 & 13 & 0.08  & -7 & 48 & 0.35 & -407 & 4 & -10 & 15 & 6 & 9.1 & 49 & -2.7$_{-1.0}^{+1.0}$ & 1 & 2  \\
 17 & 1995/04/03 07:48 & 27.0 & 10 & 0.10  & 8 & 122 & -0.47 & -302 & -25 & 46 & 16 & -19 & 12.8 & 57 & 1.5$_{-0.3}^{+0.3}$ & 2 & 2  \\
 18 & 1995/04/06 07:18 & 10.5 & 9 & 0.02  & 22 & 8 & 0.15 & -328 & -15 & 8 & -18 & -19 & 6.0 & 24 & -9.8$_{-3.5}^{+3.2}$ & 2 & 2  \\
 19 & 1995/08/22 21:18 & 22.0 & 10 & 0.08 & -36 & 232 & 0.12 & -361 & 3 & -8 & 13 & 3 & 10.4 & 61 & 1.7$_{-0.7}^{+0.7}$ & 1 & 2  \\
 20 & 1995/12/16 05:18 & 17.0 & 11 & 0.05  & 5 & 21 & 0.76 & -399 & 15 & -11 & 17 & 10 & 8.0 & 22 & -1.8$_{-0.6}^{+0.5}$ & 1 & 3  \\
 21 & 1996/05/27 15:18 & 40.0 & 19 & 0.17  & 4 & 54 & 0.59 & -358 & -22 & -10 & 14 & -7 & 19.1 & 54 & -2.6$_{-0.7}^{+0.8}$ & 2 & 2  \\
 22 & 1996/07/01 17:18 & 17.0 & 14 & 0.07  & 6 & 75 & 0.28 & -357 & -20 & -8 & 13 & 9 & 8.1 & 75 & -4.2$_{-1.2}^{+1.2}$ & 2 & 2  \\
 23 & 1996/08/07 12:18 & 22.5 & 7 & 0.09  & -40 & 299 & 0.36 & -344 & 0 & 8 & 0 & 3 & 11.3 & 67 & 1.7$_{-0.5}^{+0.5}$ & 2 & 1  \\
 24 & 1996/12/24 02:48 & 32.5 & 20 & 0.15  & 30 & 60 & -0.52 & -349 & -15 & 24 & 29 & -7 & 14.9 & 64 & 3.1$_{-0.8}^{+0.8}$ & 2 & 1  \\
 25 & 1997/01/10 05:18 & 21.0 & 37 & 0.11  & -9 & 299 & -0.49 & -433 & 6 & -17 & 6 & 0 & 10.3 & 60 & 5.4$_{-2.8}^{+3.0}$ & 2 & 1  \\
 26 & 1997/02/10 03:24 & 15.0 & 7 & 0.08  & -60 & 344 & 0.49 & -464 & 7 & 4 & 18 & 24 & 7.2 & 61 & 0.0$_{-0.2}^{+0.2}$ & 2 & 3  \\
 27 & 1997/04/11 05:36 & 13.5 & 41 & 0.07 & 24 & 176 & 0.83 & -468 & -1 & -40 & 5 & 29 & 6.7 & 24 & 6.6$_{-5.0}^{+5.1}$ & 2 & 2  \\
 28 & 1997/04/21 14:30 & 40.0 & 11 & 0.16 & 32 & 299 & 0.23 & -354 & -13 & 5 & 10 & 4 & 19.4 & 65 & 1.3$_{-1.1}^{+1.1}$ & 1 & 3  \\
 29 & 1997/05/15 09:06 & 16.0 & 19 & 0.06  & -16 & 134 & -0.05 & -453 & 33 & -4 & -7 & -5 & 8.2 & 47 & -3.3$_{-1.0}^{+1.0}$ & 2 & 2  \\
 30 & 1997/05/16 06:06 & 7.8 & 10 & 0.04 & 20 & 240 & -0.27 & -475 & -5 & 4 & 21 & 23 & 3.6 & 62 & -7.8$_{-5.1}^{+4.8}$ & 2 & 3  \\
 31 & 1997/06/09 02:18 & 21.0 & 21 & 0.06& -8 & 210 & 0.59 & -372 & -3 & 6 & 7 & 9 & 10.1 & 31 & 6.3$_{-1.7}^{+1.8}$ & 2 & 2  \\
 32 & 1997/06/19 05:06 & 10.8 & 13 & 0.05& -58 & 314 & -0.48 & -349 & -5 & 24 & 10 & 11 & 5.3 & 68 & 7.2$_{-1.4}^{+1.4}$ & 1 & 3  \\
 33 & 1997/08/03 14:06 & 11.8 & 36 & 0.04 & -2 & 36 & 0.64 & -418 & -39 & -16 & -1 & -26 & 6.0 & 36 & -11.7$_{-9.6}^{+9.3}$ & 1 & 3  \\
 34 & 1997/09/22 00:48 & 16.5 & 17 & 0.08 & 59 & 134 & 0.13 & -424 & 6 & 1 & 50 & 8 & 7.3 & 68 & -2.2$_{-0.3}^{+0.3}$ & 1 & 2  \\
 35 & 1997/10/01 16:18 & 30.5 & 10 & 0.16 & 60 & 139 & -0.24 & -441 & -11 & -3 & 9 & 8 & 14.9 & 67 & -0.7$_{-0.2}^{+0.2}$ & 2 & 2  \\
 36 & 1997/10/10 23:48 & 25.0 & 13 & 0.10& -7 & 232 & -0.17 & -400 & 12 & -5 & 25 & -11 & 11.6 & 52 & 2.5$_{-1.2}^{+1.2}$ & 1 & 1  \\
 37 & 1997/11/07 15:48 & 12.5 & 17 & 0.05& 30 & 224 & 0.10 & -417 & 14 & 18 & 7 & -9 & 6.1 & 52 & 2.7$_{-0.6}^{+0.6}$ & 1 & 2  \\
 38 & 1997/11/08 04:54 & 10.0 & 22 & 0.05& 56 & 8 & -0.61 & -366 & 3 & 4 & 8 & 19 & 4.8 & 56 & 6.0$_{-1.9}^{+2.0}$ & 1 & 2  \\
 39 & 1997/11/22 15:48 & 20.5 & 31 & 0.07  & 22 & 187 & 0.60 & -494 & 23 & -9 & 16 & 12 & 9.6 & 23 & -6.2$_{-2.1}^{+2.2}$ & 2 & 3  \\
 40 & 1998/01/07 03:18 & 29.0 & 40 & 0.14  & 56 & 134 & -0.47 & -380 & -11 & -5 & 27 & -1 & 13.4 & 67 & -4.3$_{-0.8}^{+0.8}$ & 1 & 1  \\
 41 & 1998/02/04 04:30 & 42.0 & 13 & 0.04  & 0 & 7 & -0.36 & -322 & 30 & 16 & 13 & -15 & 17.2 & 7 & -6.1$_{-2.3}^{+2.2}$ & 2 & 2  \\
\hline
\end{tabular}
\end{table*}

\begin{table*}[h]\centering
\linespread{1} \caption{(continued)}
\begin{tabular}{c|cc|cccccccccccccc|c}
\hline
 & \multicolumn{2}{c|}{MC Interval} & \multicolumn{14}{c|}{Modeled Parameters} &  \\
\cline{2-3}\cline{4-17}
No. & $t_0$ & $\Delta t$ & $B_0$ & $R$ & $\theta$ & $\phi$ & $d$ & $v_X$ & $v_Y$ & $v_Z$ & $v_{e}$ & $v_{p}$ & $\Delta t_c$ & $\Theta$ & $\tau$ & $Q$ & $Q_0$ \\
(1) &
(2) &
(3) &
(4) &
(5) &
(6) &
(7) &
(8) &
(9) &
(10) &
(11) &
(12) &
(13) &
(14) &
(15) &
(16) &
(17) &
(18) \\
\hline
 42 & 1998/03/04 14:18 & 40.0 & 26 & 0.14  & 14 & 50 & 0.44 & -338 & -5 & -3 & 9 & 0 & 19.3 & 51 & -4.4$_{-1.1}^{+1.1}$ & 2 & 1  \\
 43 & 1998/06/02 10:36 & 5.3 & 14 & 0.02 & 10 & 44 & 0.48 & -409 & -22 & -48 & 21 & 18 & 2.5 & 45 & -10.0$_{-2.4}^{+2.1}$ & 1 & 2  \\
 44 & 1998/06/24 16:48 & 29.0 & 14 & 0.15  & 45 & 120 & 0.24 & -456 & -8 & 8 & 22 & 8 & 13.8 & 69 & -1.5$_{-0.4}^{+0.4}$ & 1 & 2  \\
 45 & 1998/08/20 10:18 & 33.0 & 22 & 0.11 & -7 & 232 & 0.34 & -328 & 0 & -8 & 17 & 12 & 15.4 & 52 & 3.7$_{-1.9}^{+2.2}$ & 1 & 1  \\
 46 & 1998/09/25 10:18 & 27.0 & 23 & 0.22  & 60 & 217 & 0.55 & -627 & 86 & 57 & 84 & -19 & 11.8 & 66 & -1.7$_{-0.6}^{+0.6}$ & 1 & 2  \\
 47 & 1998/10/19 05:06 & 9.5 & 134 & 0.02 & 10 & 0 & 0.93 & -402 & 23 & -10 & 17 & -21 & 4.3 & 10 & 49.7$_{-12.1}^{+13.1}$ & 2 & 3  \\
 48 & 1998/11/08 23:48 & 25.5 & 34 & 0.13 & -45 & 159 & 0.51 & -451 & -27 & 14 & 52 & -7 & 11.1 & 48 & 4.5$_{-1.1}^{+1.1}$ & 1 & 1  \\
 49 & 1999/02/18 14:18 & 22.0 & 9 & 0.14 & -32 & 312 & -0.40 & -619 & -21 & -7 & 106 & 0 & 8.9 & 54 & -0.8$_{-0.2}^{+0.2}$ & 2 & 3  \\
 50 & 1999/04/16 20:18 & 25.0 & 24 & 0.11 & -37 & 127 & 0.10 & -412 & -19 & 1 & 30 & 3 & 11.4 & 61 & -2.8$_{-1.0}^{+1.0}$ & 2 & 3  \\
 51 & 1999/08/09 10:48 & 29.0 & 19 & 0.12 & 60 & 15 & 0.57 & -338 & 5 & -15 & -3 & 0 & 14.6 & 61 & -3.5$_{-2.2}^{+2.2}$ & 2 & 1  \\
 52 & 1999/09/21 21:06 & 8.0 & 13 & 0.03 & -4 & 132 & -0.03 & -356 & -7 & 2 & -7 & 3 & 4.2 & 47 & -7.0$_{-1.9}^{+1.9}$ & 1 & 3  \\
 53 & 2000/02/21 09:48 & 27.5 & 34 & 0.12& 54 & 330 & -0.43 & -385 & 3 & -1 & 33 & -3 & 12.4 & 59 & 4.9$_{-2.5}^{+2.8}$ & 1 & 3  \\
 54 & 2000/07/01 08:48 & 18.5 & 11 & 0.02 & 10 & 0 & 0.57 & -413 & 10 & 13 & 1 & 23 & 9.1 & 10 & -9.9$_{-3.7}^{+3.8}$ & 2 & 1  \\
 55 & 2000/07/28 21:06 & 13.0 & 14 & 0.04 & -7 & 337 & 0.50 & -463 & -22 & 8 & 11 & -4 & 6.2 & 23 & -3.2$_{-2.0}^{+1.8}$ & 2 & 2  \\
 56 & 2000/08/01 00:06 & 15.8 & 17 & 0.03 & 2 & 194 & 0.87 & -438 & -43 & 20 & 8 & -71 & 7.5 & 14 & -8.7$_{-0.9}^{+0.9}$ & 2 & 3  \\
 57 & 2000/08/12 06:06 & 23.0 & 35 & 0.13 & 7 & 52 & 0.37 & -550 & -1 & -30 & 59 & -19 & 10.0 & 52 & -2.3$_{-1.2}^{+1.2}$ & 1 & 2  \\
 58 & 2000/09/18 01:54 & 13.2 & 57 & 0.28 & 23 & 217 & -0.95 & -775 & -32 & -143 & 192 & 0 & 5.8 & 43 & -1.5$_{-1.6}^{+1.7}$ & 1 & 3  \\
 59 & 2000/10/03 17:06 & 21.0 & 20 & 0.09 & 34 & 60 & 0.18 & -399 & 6 & -14 & 15 & -18 & 10.0 & 66 & 3.1$_{-1.0}^{+1.1}$ & 1 & 1  \\
 60 & 2000/10/13 18:24 & 22.5 & 12 & 0.10 & -38 & 125 & -0.07 & -395 & -2 & -2 & 0 & 18 & 11.2 & 63 & 1.6$_{-0.8}^{+0.8}$ & 1 & 2  \\
 61 & 2000/10/28 23:18 & 25.0 & 15 & 0.11  & -23 & 119 & -0.38 & -388 & 3 & -11 & 38 & -11 & 11.1 & 63 & -1.2$_{-0.1}^{+0.1}$ & 2 & 3  \\
 62 & 2001/03/19 23:18 & 19.0 & 42 & 0.02 & 4 & 180 & -0.54 & -405 & -17 & -28 & -10 & -11 & 10.9 & 4 & 31.4$_{-12.2}^{+14.5}$ & 2 & 1  \\
 63 & 2001/03/20 17:48 & 45.0 & 15 & 0.15  & 0 & 44 & 0.54 & -325 & 10 & -32 & 54 & 0 & 18.1 & 44 & -1.7$_{-0.0}^{+0.0}$ & 2 & 3  \\
 64 & 2001/04/04 20:54 & 11.5 & 22 & 0.14 & -23 & 60 & 0.82 & -681 & -48 & -3 & 67 & -64 & 5.3 & 63 & 2.7$_{-1.3}^{+1.3}$ & 2 & 1  \\
 65 & 2001/04/12 07:54 & 10.0 & 61 & 0.08 & 8 & 198 & 0.87 & -656 & 78 & -65 & 95 & 54 & 4.2 & 20 & 8.9$_{-2.2}^{+2.4}$ & 1 & 2  \\
 66 & 2001/04/22 00:54 & 24.5 & 17 & 0.10  & -45 & 308 & 0.32 & -357 & 0 & 1 & 27 & 0 & 11.2 & 64 & -2.8$_{-0.2}^{+0.2}$ & 2 & 2  \\
 67 & 2001/04/29 01:54 & 11.0 & 45 & 0.11  & 18 & 60 & 0.73 & -636 & 14 & -39 & 39 & 1 & 5.2 & 62 & -8.6$_{-4.4}^{+4.2}$ & 2 & 2  \\
 68 & 2001/05/28 11:54 & 22.5 & 14 & 0.08  & -14 & 30 & 0.52 & -455 & 24 & 36 & -5 & 4 & 11.5 & 33 & -3.3$_{-0.8}^{+0.7}$ & 1 & 1 \\
 69 & 2001/10/31 21:18 & 37.0 & 20 & 0.14  & -7 & 119 & -0.27 & -332 & -8 & 7 & 22 & -1 & 17.1 & 60 & -4.1$_{-0.5}^{+0.5}$ & 2 & 3  \\
 70 & 2002/03/19 22:54 & 16.5 & 16 & 0.06 & 12 & 44 & -0.35 & -375 & -9 & -8 & -7 & -23 & 8.5 & 46 & 2.2$_{-0.2}^{+0.2}$ & 2 & 2  \\
 71 & 2002/03/24 03:48 & 43.0 & 28 & 0.19 & 21 & 224 & 0.47 & -435 & 0 & -4 & 5 & 6 & 21.1 & 48 & 2.6$_{-1.1}^{+1.1}$ & 2 & 2  \\
 72 & 2002/04/18 04:18 & 22.0 & 19 & 0.13 & -17 & 224 & 0.68 & -469 & -1 & 11 & 18 & -11 & 10.6 & 47 & 1.9$_{-0.0}^{+0.0}$ & 2 & 1  \\
 73 & 2002/04/20 11:48 & 29.0 & 25 & 0.21  & -14 & 53 & 0.71 & -516 & -2 & -63 & 74 & 31 & 12.8 & 54 & -3.2$_{-2.1}^{+2.1}$ & 2 & 3  \\
 74 & 2002/05/19 03:54 & 19.5 & 23 & 0.25  & -3 & 104 & 0.92 & -439 & 20 & -59 & 75 & -21 & 9.0 & 75 & -1.4$_{-0.1}^{+0.1}$ & 2 & 1  \\
 75 & 2002/05/23 23:24 & 17.5 & 12 & 0.10  & -18 & 134 & 0.78 & -560 & 300 & 22 & 5 & -98 & 8.7 & 48 & -1.2$_{-0.0}^{+0.0}$ & 2 & 3  \\
 76 & 2002/08/01 11:54 & 10.7 & 15 & 0.04 & 0 & 216 & 0.26 & -458 & 13 & 0 & 26 & 13 & 4.8 & 36 & 6.5$_{-3.3}^{+3.2}$ & 1 & 3  \\
 77 & 2002/08/02 07:24 & 13.7 & 13 & 0.07  & -11 & 240 & 0.06 & -489 & -10 & -7 & 18 & -11 & 6.5 & 61 & -1.7$_{-0.4}^{+0.4}$ & 1 & 2  \\
 78 & 2002/09/03 00:18 & 18.5 & 13 & 0.06 & 34 & 203 & 0.45 & -352 & -2 & -28 & -19 & 15 & 9.9 & 40 & 3.7$_{-2.9}^{+2.8}$ & 1 & 2  \\
 79 & 2003/06/17 17:48 & 14.5 & 16 & 0.08  & -13 & 315 & 0.63 & -490 & -4 & -23 & -35 & 25 & 7.8 & 46 & -2.9$_{-0.6}^{+0.6}$ & 1 & 3  \\
 80 & 2003/07/10 19:54 & 13.0 & 24 & 0.04 & -16 & 166 & 0.84 & -359 & -17 & -1 & -10 & 11 & 6.7 & 21 & 10.3$_{-3.8}^{+4.6}$ & 2 & 3  \\
 81 & 2003/08/18 11:36 & 16.8 & 14 & 0.09 & -60 & 314 & 0.01 & -489 & 0 & 14 & -21 & 15 & 8.8 & 69 & 1.0$_{-0.4}^{+0.5}$ & 1 & 2  \\
 82 & 2003/11/20 10:48 & 15.5 & 49 & 0.10 & -54 & 146 & 0.12 & -581 & -11 & 23 & 81 & 6 & 6.6 & 60 & 4.9$_{-1.6}^{+1.7}$ & 1 & 2  \\
 83 & 2004/04/04 02:48 & 36.0 & 17 & 0.15  & 52 & 7 & -0.13 & -434 & 17 & 3 & 12 & -14 & 17.3 & 53 & -1.2$_{-0.3}^{+0.3}$ & 2 & 2  \\
 84 & 2004/07/22 15:24 & 7.7 & 20 & 0.05 & -29 & 60 & -0.16 & -602 & 70 & -13 & -24 & -79 & 4.0 & 64 & 5.4$_{-3.1}^{+3.1}$ & 1 & 3  \\
 85 & 2004/11/08 03:24 & 13.2 & 33 & 0.04  & -2 & 15 & 0.46 & -667 & 50 & -13 & 32 & 49 & 5.7 & 15 & -12.2$_{-4.5}^{+4.1}$ & 1 & 2  \\
\hline
\end{tabular}
\end{table*}

\begin{table*}[h]\centering
\linespread{1} \caption{(continued)}
\begin{tabular}{c|cc|cccccccccccccc|c}
\hline
 & \multicolumn{2}{c|}{MC Interval} & \multicolumn{14}{c|}{Modeled Parameters} &  \\
\cline{2-3}\cline{4-17}
No. & $t_0$ & $\Delta t$ & $B_0$ & $R$ & $\theta$ & $\phi$ & $d$ & $v_X$ & $v_Y$ & $v_Z$ & $v_{e}$ & $v_{p}$ & $\Delta t_c$ & $\Theta$ & $\tau$ & $Q$ & $Q_0$ \\
(1) &
(2) &
(3) &
(4) &
(5) &
(6) &
(7) &
(8) &
(9) &
(10) &
(11) &
(12) &
(13) &
(14) &
(15) &
(16) &
(17) &
(18) \\
\hline
 86 & 2004/11/09 20:54 & 6.5 & 84 & 0.04  & 18 & 327 & 0.51 & -805 & 5 & 0 & 24 & 3 & 3.2 & 36 & -13.0$_{-2.5}^{+2.5}$ & 1 & 2  \\
 87 & 2004/11/10 03:36 & 7.5 & 38 & 0.06  & -48 & 15 & 0.59 & -725 & -46 & -12 & 67 & -10 & 3.3 & 50 & -4.5$_{-1.3}^{+1.4}$ & 1 & 2  \\
 88 & 2005/05/20 07:18 & 22.0 & 14 & 0.11  & 57 & 314 & 0.10 & -457 & -2 & -1 & 0 & 4 & 10.9 & 67 & -3.3$_{-2.0}^{+2.0}$ & 2 & 2  \\
 89 & 2005/06/12 15:36 & 15.5 & 31 & 0.06  & -22 & 164 & -0.77 & -469 & 54 & 0 & 20 & 17 & 7.2 & 27 & -6.5$_{-2.3}^{+1.9}$ & 2 & 2  \\
 90 & 2005/06/15 05:48 & 26.0 & 11 & 0.15  & 37 & 119 & -0.37 & -483 & -6 & -8 & 13 & -10 & 12.6 & 67 & -2.4$_{-0.4}^{+0.4}$ & 2 & 3  \\
 91 & 2005/07/17 15:18 & 12.5 & 14 & 0.06  & -29 & 60 & 0.16 & -426 & -8 & -24 & 32 & 1 & 5.8 & 64 & 4.3$_{-1.0}^{+1.0}$ & 1 & 2  \\
 92 & 2005/10/31 02:54 & 17.5 & 15 & 0.03  & -6 & 165 & 0.15 & -365 & -24 & 10 & 10 & -3 & 8.0 & 15 & 13.9$_{-5.1}^{+5.2}$ & 1 & 3  \\
 93 & 2005/12/31 14:48 & 20.0 & 22 & 0.02  & 1 & 356 & -0.67 & -465 & -20 & 18 & -16 & -13 & 12.1 & 3 & 27.8$_{-17.1}^{+17.6}$ & 2 & 2  \\
 94 & 2006/02/05 19:06 & 18.0 & 10 & 0.07  & -34 & 119 & 0.19 & -340 & -10 & -3 & 21 & 5 & 8.3 & 65 & 2.9$_{-2.2}^{+2.5}$ & 2 & 2  \\
 95 & 2006/04/13 14:48 & 6.0 & 19 & 0.03  & 49 & 164 & -0.25 & -531 & -2 & -2 & 6 & -13 & 3.0 & 51 & -6.5$_{-3.6}^{+3.1}$ & 1 & 3  \\
 96 & 2006/04/13 20:36 & 13.3 & 20 & 0.07  & -6 & 299 & 0.22 & -508 & 14 & 6 & 10 & -10 & 6.5 & 60 & -2.5$_{-0.9}^{+0.9}$ & 1 & 2  \\
 97 & 2006/08/30 21:06 & 17.8 & 10 & 0.08  & -6 & 299 & 0.41 & -398 & -7 & 16 & 19 & -12 & 8.5 & 60 & -3.3$_{-0.4}^{+0.4}$ & 2 & 2  \\
 98 & 2006/12/14 22:48 & 21.0 & 9 & 0.18  & -48 & 45 & -0.38 & -766 & 74 & -27 & 199 & -51 & 7.8 & 62 & -0.2$_{-0.0}^{+0.0}$ & 2 & 3  \\
 99 & 2007/01/14 14:06 & 16.8 & 14 & 0.05  & 5 & 337 & 0.46 & -350 & -78 & -6 & -20 & 0 & 9.2 & 23 & -3.7$_{-0.8}^{+0.8}$ & 1 & 3  \\
100 & 2007/03/24 03:06 & 13.8 & 15 & 0.05  & -14 & 308 & -0.31 & -360 & 7 & -17 & 16 & 0 & 6.5 & 52 & 9.6$_{-6.2}^{+6.5}$ & 1 & 3  \\
101 & 2007/05/21 22:54 & 14.7 & 13 & 0.05  & 30 & 12 & -0.13 & -453 & 24 & 20 & 18 & -7 & 6.8 & 32 & 6.4$_{-1.9}^{+2.0}$ & 2 & 2  \\
102 & 2007/12/25 15:42 & 15.1 & 3 & 0.06  & 5 & 240 & -0.33 & -347 & -7 & -3 & 13 & 4 & 7.2 & 60 & -1.1$_{-0.2}^{+0.3}$ & 2 & 3  \\
103 & 2008/12/17 03:06 & 11.3 & 11 & 0.04  & -5 & 224 & -0.61 & -337 & 0 & 7 & 19 & 2 & 5.3 & 44 & -4.6$_{-1.6}^{+1.5}$ & 2 & 2  \\
104 & 2009/01/02 06:06 & 9.0 & 6 & 0.03  & 7 & 139 & -0.24 & -409 & -41 & -10 & 0 & -6 & 4.5 & 41 & 0.9$_{-1.3}^{+1.3}$ & 1 & 3  \\
105 & 2009/02/04 00:06 & 10.8 & 14 & 0.04  & 4 & 46 & -0.58 & -366 & -10 & 6 & 19 & -5 & 5.2 & 46 & 5.6$_{-1.9}^{+2.1}$ & 1 & 2  \\
106 & 2009/03/12 00:42 & 24.0 & 14 & 0.09  & 37 & 139 & 0.39 & -361 & -17 & 0 & -27 & 6 & 13.1 & 52 & 3.6$_{-1.1}^{+1.2}$ & 2 & 2  \\
107 & 2009/06/27 15:18 & 27.0 & 12 & 0.10  & 45 & 150 & 0.18 & -385 & 9 & -1 & 10 & 7 & 13.1 & 52 & 7.3$_{-1.9}^{+1.9}$ & 2 & 3  \\
108 & 2009/07/21 03:54 & 13.2 & 10 & 0.04  & 5 & 30 & -0.80 & -318 & 0 & -8 & -1 & 6 & 6.7 & 30 & 3.7$_{-0.6}^{+0.6}$ & 2 & 2  \\
109 & 2009/09/10 10:24 & 6.0 & 7 & 0.02 & 34 & 119 & 0.50 & -306 & 3 & 0 & 5 & 2 & 3.0 & 65 & 9.1$_{-1.9}^{+1.8}$ & 1 & 2  \\
110 & 2009/09/30 07:54 & 9.0 & 9 & 0.03  & 29 & 314 & 0.25 & -348 & 6 & 18 & 10 & 5 & 4.3 & 52 & -4.4$_{-1.7}^{+1.8}$ & 1 & 2  \\
111 & 2009/10/12 12:06 & 4.8 & 7 & 0.02  & 21 & 302 & 0.26 & -363 & -9 & 2 & 0 & -14 & 2.5 & 60 & -8.0$_{-0.9}^{+1.0}$ & 2 & 2  \\
112 & 2009/10/17 22:06 & 9.3 & 8 & 0.04 & 14 & 119 & 0.71 & -316 & 3 & -13 & 13 & 3 & 4.5 & 61 & 9.8$_{-3.5}^{+3.7}$ & 1 & 3 \\
113 & 2009/10/29 05:12 & 17.6 & 10 & 0.07  & 7 & 230 & -0.46 & -367 & 8 & 18 & -20 & -5 & 9.4 & 51 & -1.8$_{-0.4}^{+0.4}$ & 2 & 3  \\
114 & 2009/11/01 08:48 & 23.0 & 21 & 0.03 & 0 & 3 & 0.91 & -343 & 24 & -12 & -6 & 32 & 12.1 & 3 & 21.4$_{-44.6}^{+44.6}$ & 2 & 2  \\
115 & 2009/12/12 19:48 & 33.5 & 8 & 0.10  & -7 & 134 & 0.59 & -266 & -8 & -6 & 13 & -1 & 15.8 & 45 & 2.3$_{-0.2}^{+0.2}$ & 2 & 3  \\
\hline
\end{tabular}
Note 1: Column 2 is the begin time of a MC in UT. Column 3 is the duration of the observed MC interval in units of hour.
The interpretations of the next thirteen columns could be found in Table~\ref{tb_par} with the difference that Column 14
is $\Delta t_c=t_c-t_0$. The values of $B_0$, $R$, $v_{p}$ and $\tau$ are all obtained at the time of $t_c$. Column 17
gives the quality of the fit and the last column is the quality of the fit given in Lepping's list.
Quality of 1 or 2 means good or fair.
One can refer to Sec.\ref{sec_model} for more details. \\ Note 2: For the modeled parameters, $B_0$ is in
units of nT, $R$ in units of AU, $\theta$, $\phi$ and $\Theta$ in units of degree, $d$ in units of $R$,
all the speeds are in units of km s$^{-1}$, $\Delta t_c$ in units of hour, $\tau$ in units of turns per AU.
\end{table*}


\end{article}
\end{document}